\documentstyle[amssymb,aps]{revtex}

\begin{document}
\title{The classical limit of non-integrable quantum systems}
\author{Mario Castagnino}
\address{CONICET-UNR-UBA, Institutos de F\'{i}sica de Rosario y de Astronom\'{i}a y\\
F\'{i}sica del Espacio.\\
Casilla de Correos 67, Sucursal 28, 1428, Buenos Aires, Argentina}
\author{Olimpia Lombardi}
\address{CONICET-IEC, Universidad Nacional de Quilmes\\
Rivadavia 2328, 6${{}^\circ}$ Derecha, Buenos Aires, Argentina.}
\maketitle

\begin{abstract}
The classical limit of non-integrable quantum systems is studied. We define
non-integrable quantum systems as those which have, as their classical
limit, a non-integrable classical system. In order to obtain this limit, the 
{\it self-induced decoherence }approach and the corresponding classical
limit are generalized from integrable to non-integrable systems. In this
approach, the lost of information, usually conceived as the result of a
coarse-graining or the trace of an environment, is produced by a particular
choice of the algebra of observables and the systematic use of mean values,
that project the unitary evolution onto an effective non-unitary one. The
decoherence times computed with this approach coincide with those of the
literature. By means of our method, we can obtain the classical limit of the
quantum state of a non-integrable system, which turns out to be a set of
unstable, potentially chaotic classical trajectories contained in the Wigner
transformation of the quantum state.

PACS number(s) 03.65.Bz

e-mail: mariocastagnino@citynet.net.ar
\end{abstract}

\section{Introduction}

The problem of the classical limit of quantum mechanics has a long history.
In the beginning, on the basis of the analogy with special relativity where
the limit $c\rightarrow \infty $ leads to the classical behavior, it was
thought that the classical limit was just the limit $\hbar \rightarrow 0$.
But it was soon realized that this was only one element of the problem,
namely, {\it macroscopicity}, and that other elements must be taken into
account: e.g. quantum mechanics has a probabilistic non-Boolean structure
while classical mechanics has a non-probabilistic and Boolean one. Thus,
necessarily two new elements must come into play:

\begin{itemize}
\item  {\it Decoherence,} that transforms the non-Boolean structure into a
Boolean one, and

\item  {\it Localization (actualization or the choice of a trajectory) }%
that, with macroscopicity -which circumvents the uncertainty principle-
turns the probabilistic structure into a non-probabilistic one.
\end{itemize}

In general, decoherence in quantum systems is defined as a process that
leads to the diagonalization of a density matrix (more precisely, to the
decay of the cross-terms in the expectation value of an observable in some
basis). In a first period, decoherence was explained as the result of the
destructive interference of the off-diagonal elements of the density matrix
(see \cite{Van Kampen}, \cite{Daneri et al.}); however, this line of
research was abandoned due to technical difficulties derived from the
formalism used to describe the process. As a consequence, decoherence begun
to be conceived as produced by the interaction between a system and its
environment. This approach gave rise to the einselection program, based on
the works of Zeh (\cite{Zeh 1970}, \cite{Zeh 1971}, \cite{Zeh 1973}) and
later developed by Zurek and coworkers (\cite{Zurek 1981}, \cite{Zurek 1982}%
, \cite{Zurek 1991}, \cite{Zurek 1993}, \cite{Zurek 1998}, \cite{Paz}, \cite
{Zurek 2003}). Although many relevant results have been obtained by means of
einselection, this approach still involves certain unsolved problems, as
those related with the explanation of the emergence of classicality in
closed quantum systems, the criterion for introducing the 'cut' between the
system and its environment, and the definition of the preferred ('pointer')
basis where the system behaves classically (see \cite{SHPMP}, \cite{Max-1}).
As the result of these and other difficulties, a number of alternative
accounts of decoherence have been proposed (see \cite{Casati-1}, \cite
{Casati-2}, \cite{Penrose}, \cite{Diosi}, \cite{Milburn}, \cite{Adler}).

In a series of papers (\cite{Cast-Gadella 1997}, \cite{Cast-Laura 1997}, 
\cite{Laura-Cast 1998-E}, \cite{Laura-Cast 1998-A}, \cite{Cast 1999}, \cite
{Cast-Laura 2000-PRA}, \cite{Cast-Laura 2000-IJTP}, \cite
{Cast-Gad-Laura-Betan 2001-PLA}, \cite{Cast-Gad-Laura-Betan 2001-JPA}, \cite
{Cast-Lombardi 2003}, \cite{Cast-Gadella 2003}, \cite{Cast-Ordoñez 2004}, 
\cite{Cast 2004}, \cite{Cast-Lombardi 2005-PS}, \cite{SHPMP}) we have
returned to the initial idea of the destructive interference of the
off-diagonal terms of the density matrix, but now on the basis of a
different formalism: the formalism introduced by van Hove (\cite{van Hove
1955}, \cite{van Hove 1956}, \cite{van Hove 1957}, \cite{van Hove 1959}, 
\cite{van Hove 1979}). We have called this new approach '{\it self-induced
decoherence}' \cite{SHPMP} because, from this viewpoint, decoherence is not
produced by the interaction between a system and its environment, but
results from the own dynamics of the whole quantum system governed by a
Hamiltonian with continuous spectrum. In this approach, the difficulties
derived from the einselection program are absent: self-induced decoherence
can be used in closed systems as the universe \cite{Cast-Lombardi 2003}, the
definition of a convenient subalgebra plays the role of the coarse-graining
induced by the environment, avoiding the 'cut' problem \cite{SHPMP}, and the
pointer basis is perfectly defined (see \cite{SHPMP}, \cite{Cast-Lombardi
2005-PS} and Section III.C below).

Self-induced decoherence is capable of addressing relevant problems from a
general perspective, e.g. the problem of supplying a good definition of the
classical limit {\it in all cases}\footnote{%
Precisely: in all cases where the system do have a classical limit; e.g.,
systems with no quasi-continuous limit, yielding to a non-continuous energy
spectrum, are excluded \cite{Max-2}.}. Let us explain the essence of the
idea supporting this new approach. When we deal with continuous spectra, the
destructive interference is embodied in the Riemann-Lebesgue theorem which
states that, if $f(\nu )\in {\Bbb L}_{1}$, then 
\[
\lim_{t\rightarrow \infty }\int_{-a}^{a}f(\nu )e^{-i\frac{\nu t}{\hbar }%
}d\nu =0 
\]
where $e^{-i\frac{\nu t}{\hbar }}$ is the $\nu -$oscillating factor that
produces the destructive interference. In the case of decoherence, $\nu
=\omega -\omega ^{\prime }$, where $\omega ,\omega ^{\prime }$ are the
continuous indices of the density operator $\rho (\omega ,\omega ^{\prime })$
in the energy eigenbasis; then, $\nu =0$ corresponds to the diagonal.
However, to require that $f(\nu )\in {\Bbb L}_{1}$ is to ask too much
regularity to function $f(\nu )$, because in this case not only the
off-diagonal ($\nu \neq 0$) terms, but also the diagonal ($\nu =0)$ ones
will vanish when $t\rightarrow \infty $. Therefore, we use less regular
functions, precisely $f(\nu )=A\delta (\nu )+f_{1}(\nu )$, where only $%
f_{1}(\nu )\in {\Bbb L}_{1}$. In this case, 
\[
\lim_{t\rightarrow \infty }\int_{-a}^{a}f(\nu )e^{-i\frac{\nu t}{\hbar }%
}d\nu =A 
\]
and the diagonal terms $A\delta (\nu )$ remain while the off-diagonal terms $%
f_{1}(\nu )$ vanish. We will apply this main idea to the {\it non-integrable
case}, and present the computations in all detail in Section III.B, by using
our previous results on quantum systems with continuous spectrum contained
in papers \cite{Cast-Laura 1997}, \cite{Laura-Cast 1998-E}, \cite{Laura-Cast
1998-A}, \cite{Cast-Laura 2000-IJTP}, \cite{Cast-Gad-Laura-Betan 2001-PLA}
and \cite{Cast-Gad-Laura-Betan 2001-JPA}. With this strategy we have already
obtained, in paper \cite{Cast-Laura 2000-PRA}, the classical limit for {\it %
integrable} systems. We have also presented this result in more rigorous
mathematical basis in \cite{Cast 2004} and explained the physical
foundations of the method in papers \cite{SHPMP} and \cite{Cast-Lombardi
2005-PS}. The mathematical basis of the theory is explained in papers \cite
{Antoniou} and \cite{Cast-Ordoñez 2004}. But, of course, the big challenge
to prove the consistency and generality of the method is to find its version
for {\it non-integrable} systems, obtaining unstable, potentially chaotic
classical trajectories as a final result, which could explain models as
those of ref. \cite{Casati-Prosen}.

In the case of integrable systems, the classical limit was obtained by a
combination of the van Hove formalism and the Weyl-Wigner-Moyal isomorphism
in a globally defined pointer basis. But in the non-integrable case, such a
global basis does not exist. Nevertheless, the just quoted isomorphism is
what allows us to relax the global condition and to generalize the
formalism: quantum mechanics is formulated in a phase space that is covered
with charts where {\it local pointer bases can be defined}. The set of all
these local pointer bases will yield decomposition (\ref{A1.1}), which is
the essential tool of this paper.

The formalism of the theory is presented in a self-comprehensive way, with a
mathematics as simple as possible and in the simplest possible case; this
seems enough for the physical purposes of this paper. In Section II, a brief
review of the Weyl-Wigner-Moyal mapping is developed, and in Section III,
the theory of decoherence in non-integrable systems is explained. In Section
IV, the classical limit of quantum non-integrable system is obtained. In
Section V, the localization phenomena is briefly discussed, and in Section
VI, our previous results are generalized to the case of partially
non-integrable systems. In the conclusion (Section VII), we list the
possible future applications of the theory and explain why it could be
considered as a {\it minimal formalism for quantum chaos}. Finally, in
Appendix A we explain the integration of two systems of differential
equations relevant to our theory, and in Appendix B we give an example of
non-integrable system.

\section{Weyl-Wigner-Moyal mapping}

Let ${\cal M=M}_{2(N+1)}\equiv {\Bbb R}^{2(N+1)}$ be the phase space of our
classical system$._{\text{ }}$The functions over this phase space will be
called $f(\phi )$, where $\phi $ symbolizes the coordinates over ${\cal M}$%
\begin{equation}
\phi ^{a}=(q^{1},...,q^{N+1},p_{q}^{1},...,p_{q}^{N+1})\qquad a=1,2,...2(N+1)
\label{L.2.3}
\end{equation}
As it is known (see \cite{Wigner}, \cite{Symb}), we can map $\widehat{{\cal A%
}},$ the algebra of regular operators $\widehat{O}$ of our quantum system,
on ${\cal A}_{q},$ the algebra of integrable functions over ${\cal M}$, via
the {\it Wigner symbol} 
\begin{equation}
symb:\widehat{{\cal A}}\rightarrow {\cal A}_{q}\qquad symb\widehat{O}=O(\phi
)  \label{L.2.4}
\end{equation}
Precisely: let us consider that ${\cal M}$ has a symplectic form 
\begin{equation}
\omega _{ab}=\left( 
\begin{array}{ll}
\text{ }0 & I_{N+1} \\ 
-I_{N+1} & 0
\end{array}
\right) \qquad \omega ^{ab}=\left( 
\begin{array}{ll}
0 & -I_{N+1} \\ 
I_{N+1} & \text{ }0
\end{array}
\right)  \label{L.2.5}
\end{equation}
Then,

\begin{equation}
symb\widehat{f}\circeq f(\phi )=\int d^{2(N+1)}\psi \exp \left( \frac{i}{%
\hbar }\psi ^{a}\omega _{ab}\psi ^{b}\right) Tr\left( \widehat{T}(\psi )%
\widehat{f}\right)  \label{L.2.6}
\end{equation}
where $\widehat{f}\in \widehat{{\cal A}}$, $f(\phi )\in {\cal A}_{q}$, and 
\begin{equation}
\widehat{T}(\psi )=\exp \left( \frac{i}{\hbar }\psi ^{a}\omega _{ab}\widehat{%
\phi }^{b}\right)  \label{L.2.7}
\end{equation}
On ${\cal A}_{q}$ we can define the {\it star product} (i.e. the classical
operator related with the multiplication on $\widehat{{\cal A}}$ and,
therefore, defining the corresponding operation on ${\cal A}_{q}$) as 
\begin{equation}
symb(\widehat{f}\widehat{g})=symb\widehat{f}*symb\widehat{g}=(f*g)(\phi )
\label{L.2.8}
\end{equation}
It can be proved (\cite{Wigner}, eq.(2.59)) that 
\begin{equation}
(f*g)(\phi )=f(\phi )\exp \left( -\frac{i\hbar }{2}\overleftarrow{\partial }%
_{a}\omega ^{ab}\overrightarrow{\partial }_{b}\right) g(\phi )=g(\phi )\exp
\left( \frac{i\hbar }{2}\overleftarrow{\partial }_{a}\omega ^{ab}%
\overrightarrow{\partial }_{b}\right) f(\phi )  \label{L.2.9}
\end{equation}
We also define the {\it Moyal bracket} as the symbol corresponding to the
commutator in $\widehat{{\cal A}}$%
\begin{equation}
\{f,g\}_{mb}=\frac{1}{i\hbar }(f*g-g*f)=symb\left( \frac{1}{i\hbar }%
[f,g]\right) =\frac{1}{i\hbar }f(\phi )\sin \left( -\frac{i\hbar }{2}%
\overleftarrow{\partial }_{a}\omega ^{ab}\overrightarrow{\partial }%
_{b}\right) g(\phi )  \label{L.2.10}
\end{equation}
In the limit $\hbar \rightarrow 0$, the star product becomes the ordinary
product, and the Moyal bracket becomes the Poisson bracket\footnote{%
From eq. (\ref{L.2.9}) it is clear that these $0(\hbar )$ are continuous
functions in the limit $\hbar =0.$ This fact will be important in Section V.}
\begin{equation}
(f*g)(\phi )=f(\phi )g(\phi )+0(\hbar )  \label{L.2.11}
\end{equation}
\begin{equation}
\{f,g\}_{mb}=\{f,g\}_{pb}+0(\hbar ^{2})  \label{L.2.12}
\end{equation}
Then, we can either say that when $\hbar \rightarrow 0$ the quantum
structure tends to the classical one, or that when $\hbar \neq 0$ the
classical structure is {\it quantized} or {\it deformed} into the quantum
one.

Let us observe that if $\widehat{f}$ commutes with $\widehat{g}$, eqs.(\ref
{L.2.9}) and (\ref{L.2.11}) change to 
\begin{equation}
(f*g)(\phi )=f(\phi )\cos \left( -\frac{i\hbar }{2}\overleftarrow{\partial }%
_{a}\omega ^{ab}\overrightarrow{\partial }_{b}\right) g(\phi )
\label{L.2.13}
\end{equation}

\begin{equation}
(f*g)(\phi )=f(\phi )g(\phi )+0(\hbar ^{2})  \label{L.2.14}
\end{equation}
as it can be proved using eq.(\ref{L.2.9}).

Finally, if we want that the mapping $symb$ be one-to-one, we must define a
unique inverse of $symb$, namely, the usual quantization rule $q\rightarrow 
\widehat{q},$ $p\rightarrow \widehat{p}$ endowed with a unique ordering
prescription, e.g. the symmetrical or {\it Weyl ordering prescription }that
maps 
\begin{equation}
symb^{-1}(qp)=\frac{1}{2}\left( \widehat{q}\widehat{p}+\widehat{p}\widehat{q}%
\right)  \label{L.2.24}
\end{equation}
Then, we have 
\begin{equation}
symb^{-1}:{\cal A}_{q}{\cal \rightarrow }\widehat{{\cal A}},\quad symb:%
\widehat{{\cal A}}{\cal \rightarrow A}_{q}  \label{L.2.25}
\end{equation}
The one-to-one mapping so defined is the {\it Weyl-Wigner-Moyal symbol}.
With $symb^{-1.}$ we can 'deform' the classical system and obtain a quantum
mechanical system. With $symb$ we go from usual quantum mechanics to a
quantum mechanics 'alla classica', formulated over a phase space ${\cal M}$,
that will become the usual classical picture in the limit $\hbar \rightarrow
0$ (as we will explain below in detail). The relation between the two
structures, given by eq.(\ref{L.2.25}) (and eq.(\ref{L.2.23}) below), is an
isomorphism that we will call {\it Weyl-Wigner-Moyal isomorphism}, the only
one we will use in this paper.

Since $\widehat{{\cal A}}$ is a space of operators on a Hilbert space ${\cal %
H}$, so it is its dual $\widehat{{\cal A}^{\prime }}$; then, as it is known,
the symbol for any $\widehat{\rho }\in \widehat{{\cal A}^{\prime }}$ is
defined as\footnote{%
In the case of states, we must add a new factor $(2\pi \hbar )^{-(N+1)}$ to
definition (\ref{L.2.6}) in order to preserve the usual normalization of $%
\rho (\phi ).$ However, $\rho (\phi )$ is not non-negatively defined. With
decoherence and $\hbar \rightarrow 0$ we will obtain a non-negatively
defined $\rho (\phi )$, and ${\cal A}_{q}\rightarrow {\cal A}$, the
classical boolean algebra of ${\Bbb L}_{1}$ operators over ${\cal M}$.} 
\begin{equation}
\rho (\phi )=symb\widehat{\rho }=(2\pi \hbar )^{-(N+1)}symb_{\text{(for
operators)}}\widehat{\rho }  \label{L.3.7}
\end{equation}
where the $symb$ for operators is defined by the eqs.(\ref{L.2.6}) and (\ref
{L.2.7}). From this definition, we have (see \cite{Wigner}, eq.(2.13)) 
\begin{equation}
(\widehat{\rho }|\widehat{O})=(symb\widehat{\rho }|symb\widehat{O})=\int
d\phi ^{2(N+1)}\rho (\phi )O(\phi )  \label{L.3.8}
\end{equation}
and in $\widehat{{\cal A}}$ and $\widehat{\text{ }{\cal A}^{\prime }}$ all
the equations are the usual ones (i.e. those of papers \cite{Wigner} and 
\cite{Symb}). Let us remark that the last equation is the cornerstone of our
theory of the classical limit. In fact, as we will see, {\it it will remain
the same when we go from regular to singular objects}. Once this statement
is understood, the translation from the quantum language to the classical
one will be easy.

.

\section{Decoherence in non-integrable systems}

\subsection{Local CSCO}

a.- We will begin with demonstrating an important theorem: when our quantum
system is endowed with a CSCO of $N+1$ observables containing $\widehat{H}$
that defines a basis in terms of which the state of the system can be
expressed, the underlying classical system is {\it integrable}. In fact, let
a classical system be defined in a phase space ${\cal M\equiv }{\Bbb R}%
^{2(N+1)}$ that can be deformed 'alla Weyl'. If our quantum system is
endowed with a $N+1-$CSCO $\{\widehat{H,}\widehat{O}_{1},...,\widehat{O}$ $%
_{N}\}$, the Moyal brackets of these quantities are 
\begin{equation}
\{O_{I}(\phi ),O_{J}(\phi )\}_{mb}=symb\left( \frac{1}{i\hbar }[\widehat{O}%
_{I},\widehat{O}_{J}]\right) =0  \label{l.1}
\end{equation}
where $I,$ $J,...=0,1,...,N$ and $\widehat{H}=\widehat{O}_{0}.$ Then, when $%
\hbar \rightarrow 0$, from eq.(\ref{L.2.12}) we know that 
\begin{equation}
\{O_{I}(\phi ),O_{J}(\phi )\}_{pb}=0  \label{l.2}
\end{equation}
Thus, as $H(\phi )=O_{0}(\phi ),$ the set $\{O_{I}(\phi )\}$ is a complete
set of $N+1$ constants of the motion in involution, globally defined over
all ${\cal M}$ and, therefore, the system is integrable. q. e. d.

As a consequence, {\it non-integrable} classical systems (precisely
classical and also macroscopic ones such that $\hbar \approx 0$), in their
quantum version, cannot have a CSCO of $N+1$ observables globally defined
containing $\widehat{H}$. But, according to the self-induced approach, the
pointer basis is precisely the eigenbasis a global $N+1-$CSCO (in such a way
that the vectors of the pointer basis turn out to be stationary states, see 
\cite{Cast-Laura 2000-PRA}). Therefore, pointer bases cannot be globally
defined in non-integrable systems. These systems can be adequately
quantized, but it is impossible (at least globally) to define a complete
stationary eigenbasis of $N+1-$CSCO and, a fortiori, a pointer $N+1-$CSCO or
a pointer basis where the system would decohere according to the
self-induced approach.\footnote{%
Observe that, if the CSCO has $<N+1$ operators, we have not {\it good}
quantum numbers enough to label the eigenvectors.} This is the main problem
with non-integrable quantum systems.\footnote{%
In the 'old quantization' approach, the problems were certainly more severe.}

b.- We will now prove that $N+1$ constants of the motion in involution
always exist locally.\footnote{%
This fact can be considered as almost evident, but since it is not
demonstrated in usual textbooks, we will give a complete demonstration below.%
} Let us consider a non-integrable quantum system (i.e. with no global $N+1-$%
CSCO), but let us suppose that, as usual, $H(\phi )=symb\widehat{H}$ is
globally defined over ${\cal M}$ (this means that any non-global CSCO has at
least one global observable: $\widehat{H}$).\footnote{%
We can also say that the quantum system is {\it dissymmetrized} \cite{Naka}.}
Now we can try to find $N$ constants of the motion $\{O_{I}(\phi )\}$ ($%
I=1,2,...,N)$ satisfying 
\begin{equation}
\{H(\phi ),O_{I}(\phi )\}_{pb}=\sum_{j=1}^{N}\frac{\partial H}{\partial
p_{qj}}\frac{\partial O_{I}}{\partial q_{j}}-\frac{\partial H}{\partial q_{j}%
}\frac{\partial O_{I}}{\partial p_{qj}}=0  \label{1.3}
\end{equation}
This is a system of $N$ partial differential equations which, with adequate
boundary conditions, has a unique solution in a {\it maximal domain of
integration }${\cal D}_{\phi _{i}}${\it \ around any point }$\phi _{i}\in 
{\cal M}$ (provided that the functions involved satisfy reasonable -e.g.
Lipschitzian- mathematical conditions that we assume\footnote{%
See Appendix A for details. Moreover, a certain determinant $\Delta $,
defined in \cite{CH}, must be $\Delta \neq 0$ in this domain.}).

But we would like to obtain a set of constants of the motion in {\it %
involution}. Then, let us suppose that $N$ different initial conditions for
eq.(\ref{1.3}) are given in a $2N+1$ dimensional hypersurface containing $%
\phi _{i}$, that we will call ${\cal D}_{\phi _{i}}^{N}$. Integrating (\ref
{1.3}) we will obtain $N$ constants of the motion $O_{I}(\phi )$. Moreover,
we can easily show that, if these solutions are in involution in ${\cal D}%
_{\phi _{i}}^{N}$, they will remain in involution in the domain ${\cal D}%
_{\phi _{i}}={\cal D}_{\phi _{i}}^{N+1}$ of $2(N+1)$ dimensions. In fact,
according to the Jacobi property of the Poisson brackets we have: 
\begin{equation}
\{H(\phi ),\{O_{I}(\phi ),O_{J}(\phi )\}_{pb}\}_{pb}+\{O_{I}(\phi
),\{O_{J}(\phi ),H(\phi )\}_{pb}\}_{pb}+\{O_{J}(\phi ),\{H(\phi ),O_{I}(\phi
)\}_{pb}\}_{pb}=0  \label{l.4}
\end{equation}
Then, since $O_{I}(\phi )$ and $O_{J}(\phi )$ are constants of the motion in 
${\cal D}_{\phi _{i}}$, the $\{O_{I}(\phi ),O_{J}(\phi )\}_{pb}$ will also
be so. As a consequence, if we could define $N$ constants of the motion such
that 
\begin{equation}
\{O_{I}(\phi ),O_{J}(\phi )\}_{pb}=0  \label{1.5}
\end{equation}
at each point $\phi \in {\cal D}_{\phi _{i}}^{N}$ (where ${\cal D}_{\phi
_{i}}^{N}$ is the already defined domain of $2N+1$ dimensions around $\phi
_{i}$) using these functions as initial conditions, we can obtain a complete
set of constants of the motion in involution in the domain ${\cal D}_{\phi
_{i}}={\cal D}_{\phi _{i}}^{N+1}$ of dimension $2(N+1)$, as promised.

Now the problem is reduced to prove the existence of the $N$ $O_{I}(\phi ),$ 
$O_{J}(\phi )$ satisfying eq.(\ref{1.5}) in ${\cal D}_{\phi _{i}}^{N}.$
Again, the existence of such a set can be easily proved by using the same
strategy as above, but now {\it recursively}. We can begin with an arbitrary
function $O_{1}(\phi )$ defined in a domain ${\cal D}_{\phi _{i}}^{0}$ of $%
N+1$ dimensions. Then, we consider another $O_{2}(\phi )$ (defined in a $N+2$
dimension domain ${\cal D}_{\phi _{i}}^{1}$ containing ${\cal D}_{\phi
_{i}}^{0}$) as the Hamiltonian of eq.(\ref{1.3}) and obtain by integration a
function $O_{1}(\phi )$, defined in the domain ${\cal D}_{\phi _{i}}^{1}$ of 
$N+2$ dimensions, such that in this domain $\{O_{2}(\phi ),O_{1}(\phi
)\}_{pb}=0$. Finally, we iterate the procedure up to find the set of
functions in involution in the ${\cal D}_{\phi _{i}}^{N}$ of dimensions $%
2N+1 $, which can be taken as initial conditions of eq.(\ref{1.3}). In this
way, the proof is completed.

c.- Now, in order to go from classical to quantum, we can also extend these
local $O_{I}(\phi )$, defined in ${\cal D}_{\phi i}={\cal D}_{\phi i}^{N+1}$
of dimensions $2(N+1)$, to all ${\cal M}$ by defining $O_{I}(\phi )=0$ for $%
\phi \in {\cal M\backprime }{\cal D}_{\phi _{i}}$. In this case, there will
be a jump in the frontier of ${\cal D}_{\phi _{i}}$, and the definition will
be only continuous a.e. (almost everywhere). Or, on physical grounds, we can
take the precaution of joining these zero functions with functions $%
O_{I}(\phi )$ in a zone around ${\cal D}_{\phi _{i}},$ that we will call $%
{\cal F}_{\phi _{i}},$ in an smooth way (e.g. by using $C^{r}$ functions
with an adequate $r$).

Therefore, we have proved the existence of local complete systems of
constants of the motion in involution $\{O_{I}(\phi )\}=\{H(\phi
),O_{1}(\phi ),...,O_{N}(\phi )\}$ that we can extend to all ${\cal M}$, at
least a.e., by adding null functions in ${\cal M}\backprime $ ${\cal D}%
_{\phi _{0}}^{N+1}$ as explained above. Since they belong to ${\cal D}_{\phi
_{i}}$, we will call them $\{H(\phi ),O_{\phi _{i}1}(\phi ),...,O_{\phi
_{i}N}(\phi )\}$. Each system $\{H(\phi ),O_{\phi _{i}1}(\phi ),...,O_{\phi
_{i}N}(\phi )\}$ can be considered as a {\it local (approximate) } $N+1-$%
CSCO in ${\cal D}_{\phi _{i}}={\cal D}_{\phi _{i}}^{N+1}$ in the sense that,
even if it is not an exact CSCO, we can compute their Weyl transformations
obtaining 
\[
\{\widehat{H}_{\phi _{i}},\widehat{O}_{\phi _{i}1},...,\widehat{O}_{\phi
_{i}N}\} 
\]
and their Wigner transformations are a complete set of constants of the
motion in involution in ${\cal D}_{\phi _{i}}.$ In fact, from eq.(\ref
{L.2.12}) we see that 
\begin{equation}
\{O_{\phi _{i}I}(\phi ),O_{\phi _{j}J}(\phi )\}_{mb}=0(\hbar ^{2}),\text{
\quad or }[\widehat{O}_{\phi _{i}I},\widehat{O}_{\phi _{j}J}]=0(\hbar ^{2})
\label{M.1}
\end{equation}
namely, they only commute approximately.

Let us now consider in more detail the joining zones ${\cal F}_{\phi _{i}}$
where we have used $C^{r}$-functions that do not satisfy the required $_{{}}$%
differential equations (\ref{1.3}) to (\ref{1.5}), in such a way that the
terms $\frac{i\hbar }{2}\overleftarrow{\partial }_{a}\omega ^{ab}%
\overrightarrow{\partial }_{b}$ of eq.(\ref{L.2.10}) produce unwanted
contributions of order $\hbar /PQ$, where $P$ and $Q$ are of the order of
magnitude of the jumps in the momentum and configuration variables in the
joining zone. Since $PQ=\varepsilon ^{2}$ is an action measuring the joining
zone (where $\varepsilon $ is the characteristic mean width of the joining
zone, precisely $\varepsilon ^{2(N+1)}\cong V_{\varepsilon }$, the volume of
the joining zones ${\cal F}_{\phi _{i}}$), the unwanted terms are of the
order of $\hbar /\varepsilon ^{2}$, that is, they are another contribution $%
0(\frac{\hbar ^{2}}{\varepsilon ^{4}})$, or simply $0(\hbar ^{2})$, to add
to (\ref{M.1})\footnote{%
Any counterfactual $0(\hbar )$ is really a factual $0(\hbar /S)$, where the
action $S\rightarrow \infty $. Precisely, this means that $S>>S_{0}$, where $%
S_{0}$ is a characteristic action. In order to give an example for this $%
S_{0}$, we can consider the fine structure constant: 
\[
\alpha =\frac{\hbar }{m_{e}ca_{0}}=\frac{\hbar }{S_{0}}\approx \frac{1}{137} 
\]
where $m_{e}$ is the electron mass, $c$ is the velocity of light, and $a_{0}$
is the Bohr radius. For $S\gg S_{0}$ we can neglect $\alpha $ and,
therefore, we loose pure quantum effects, like spin, and pass from the realm
of quantum mechanics to the classical limit.}. Anyhow, these terms will
vanish when we make the limit $\hbar \rightarrow 0$ in Section IV.

Although this approximation seems sufficient for physical purposes, we can
even improve it. In fact, we can repeat all the process based on eqs.(\ref
{1.3}) and (\ref{1.5}) substituting them with 
\begin{equation}
\{H(\phi ),O_{I}(\phi )\}_{mb}=0,\text{ }\{O_{I}(\phi ),O_{J}(\phi )\}_{mb}=0
\label{M.2}
\end{equation}
Taking into account eqs.(\ref{L.2.9}) and (\ref{L.2.12}), these are
differential equations of infinite order (see Appendix A.c). But we can cut
these expansions at any finite order and obtain a system of usual finite
differential equations that can be solved; then, we can repeat the procedure
described above with any desired precision. In this way, we can eliminate
the $0(\hbar ^{2})$ coming from eq.(\ref{L.2.12}) but not those coming from
the joining zones. Thus, we have again obtained approximate (up to $0(\hbar
^{2})$) {\it local} $N+1-$CSCO,\footnote{%
Counterfactually, $0(\frac{\hbar ^{2}}{\varepsilon ^{4}})$ goes to zero when 
$\hbar \rightarrow 0$. Factually, it goes to zero if the action of the
system $S$ is infinitely large. Then, since $\varepsilon ^{2}$ has the only
constraint $\varepsilon ^{2}<S$, it can be as large as we wish and we have $%
\frac{\hbar ^{2}}{\varepsilon ^{4}}\rightarrow 0$ (see also footnote 13).}
and we can define local eigenstates and write equations like (\ref{1.6}) for
each ${\cal D}_{\phi _{i}}$.\footnote{%
An example of this phenomenon is the Sinai billiard discussed in Appendix B.
Other examples are classical scattering systems: in fact, they have an 'in'
CSCO and an 'out' CSCO, which are different since the constants of the
motion are not the same in these CSCOs. Another example is the two slits
experiment when we mimic the screen with an infinite potential wall: before
the screen we have a local CSCO $\{\widehat{H},\widehat{P}\}$, on the screen
the CSCO is \{$\widehat{H},\widehat{X}\}$ since the kinetic term of the
Hamiltonian can be neglected with respect to the infinite potential wall,
and after the screen again $\{\widehat{H},\widehat{P}\}$. More complex
examples are the so-called pseudointegrable systems (\cite{Stockmann}, p.98, 
\cite{Ric81}, \cite{Shi95}). Tori become spheres with 'handles' that cannot
be covered with a single chart. A further example is Robnik's billiard \cite
{Berry}. Moreover, it is clear that the fractal structure of some examples
of chaos breaks the tori completely and, therefore, in this case the radius
of the integration domains probably vanish (nevertheless, this would not be
a physical case, see Appendix A).}

d.- Let us observe that natural global coordinates $\phi =(q,p_{p})$ of
phase space ${\cal M}$ can be (locally) substituted, by using (local)
canonical transformation, with (local) coordinates $(\theta _{_{\phi
_{i}}I},O_{\phi _{i}I})$, with $i=0,1,...N$ and $H=O_{\phi _{i}0}$, where
the $\theta _{_{\phi _{i}}I}(\phi )$ are the coordinates canonically
conjugated to the $O_{\phi _{i}I}(\phi )$ in ${\cal D}_{\phi _{i}}$. The $%
(\theta _{_{\phi _{i}}I},O_{\phi _{i}I})$ is clearly a chart of ${\cal M}$
in the domain ${\cal D}_{\phi _{i}}$.\footnote{%
This is not a generic chart, but a very peculiar one, since coordinates $%
O_{\phi _{i}I}$ are constants of the motion satisfying eqs.(\ref{1.3}) and (%
\ref{1.5}).} Since the system is endowed with adequate smooth properties
(let us say $C^{r}$), another similarly constructed chart $(\theta _{_{\phi
J}I},O_{\phi _{J}I})$ in the domain ${\cal D}_{\phi _{J}}$ is smoothly
connected with the previous one at any $\phi \in $ ${\cal D}_{\phi _{i}}$ $%
\cap $ ${\cal D}_{\phi _{j}}$ (see demonstration in Section V). Then, the
set of all these charts is a $C^{r}-$atlas in ${\cal M}$. This will be the
atlas we will primarily concerned with.

e.- We can also define a ({\it ad hoc}){\it \ positive partition of the
identity (}see \cite{Benatti} sec. 3.4) in the following sense. Let us
define 
\begin{equation}
1=I(\phi )=\sum_{i}B_{\phi _{i}}(\phi )  \label{PI.1}
\end{equation}
where $B_{\phi _{i}}(\phi )$ are 'bump' functions such that 
\begin{equation}
B_{\phi _{i}}(\phi )\left\{ 
\begin{array}{l}
=1\text{ if }\phi \in D_{\phi _{i}} \\ 
\in [0,1]\text{ if }\phi \in F_{\phi _{i}} \\ 
=0\text{ if }\phi \notin D_{\phi _{i}}\cup F_{\phi _{i}}
\end{array}
\right.  \label{PI.2}
\end{equation}
where $D_{\phi _{i}}$ is a domain and $F_{\phi _{i}}$ is the frontier zone
around $D_{\phi _{i}}$ (the $F_{\phi _{i}}$ are similar to the ${\cal F}%
_{\phi _{i}}$ but they are related to the $D_{\phi _{i}}$) defined in such a
way that $D_{\phi _{i}}\cup F_{\phi _{i}}\subset {\cal D}_{\phi _{i}}$ and
the intersection zones of the $D^{\prime }s$ vanish: $D_{\phi _{i}}$ $\cap $ 
$D_{\phi _{j}}=\emptyset $. Let us stress that the $B_{\phi _{i}}(\phi )$ in
the frontier zones satisfy eq.(\ref{PI.1}). Now, for any $A(\phi )$ we can
define a 
\[
A_{\phi _{i}}(\phi )=A(\phi )B_{\phi _{i}}(\phi ) 
\]
and for any $A(\phi )$ we have 
\[
A(\phi )=A(\phi )\sum_{i}B_{\phi _{i}}(\phi )=\sum_{i}A_{\phi _{i}}(\phi ) 
\]
With the mapping $symb^{-1}$ we find 
\begin{equation}
\widehat{A}=symb^{-1}A(\phi )=\sum_{i}symb^{-1}A_{\phi _{i}}(\phi )=\sum_{i}%
\widehat{A}_{\phi _{i}}  \label{PI.3}
\end{equation}
where $\widehat{A}_{\phi _{i}}=symb^{-1}A_{\phi _{i}}(\phi )$ can be
considered as a {\it localization }of $\widehat{A}$ in $D_{\phi _{i}}$.
Then, from eq.(\ref{PI.3}) 
\begin{equation}
\widehat{A}=\sum_{i}\widehat{A}_{\phi _{i}}  \label{PI.6}
\end{equation}
Moreover, since we have a local $N+1-$CSCO in each $D_{\phi _{i}}\cup
F_{\phi _{i}}\subset {\cal D}_{\phi _{i}}$, we can decompose 
\begin{equation}
\widehat{A}_{\phi _{i}}=\sum_{j}A_{j\phi _{i}}|j\rangle _{\phi
_{i}}^{(A)}\langle j|_{\phi _{i}}^{(A)}  \label{PI.8}
\end{equation}
where the \{$|j\rangle _{\phi _{i}}^{(A)}\}$ are the corresponding
eigenvectors of $\widehat{A}_{\phi _{i}}$; the $\left\{ \widehat{A}_{\phi
_{i}}\right\} $ can be considered as a local $N+1-$CSCO of $D_{\phi _{i}}$ $%
\subset {\cal D}_{\phi _{i}}.$

Now we can prove that the support of $symb|j\rangle _{\phi
_{i}}^{(A)}\langle j|_{\phi _{i}}^{(A)}$ is contained in $D_{\phi _{i}}\cup
F_{\phi _{i}}$ i.e. the support of $symb\widehat{A}_{\phi _{i}}$. In fact,
from eq.(\ref{PI.8}) we have 
\[
\widehat{A}_{\phi _{i}}|j\rangle _{\phi _{i}}^{(A)}=A_{j\phi _{i}}|j\rangle
_{\phi _{i}}^{(A)} 
\]
or 
\[
\widehat{A}_{\phi _{i}}|j\rangle _{\phi _{i}}^{(A)}\langle j|_{\phi
_{i}}^{(A)}=A_{j\phi _{i}}|j\rangle _{\phi _{i}}^{(A)}\langle j|_{\phi
_{i}}^{(A)} 
\]
Then, 
\begin{equation}
symb\widehat{A}_{\phi _{i}}*symb|j\rangle _{\phi _{i}}^{(A)}\langle j|_{\phi
_{i}}^{(A)}=A_{j\phi _{i}}symb|j\rangle _{\phi _{i}}^{(A)}\langle j|_{\phi
_{i}}^{(A)}  \label{29'}
\end{equation}
But $symb\widehat{A}_{\phi _{i}}$ and all its derivatives vanish for $\phi
\notin D_{\phi _{i}}\cup F_{\phi _{i}}$. Therefore, if $A_{j\phi _{i}}\neq 0$%
, this also must happen for $symb|j\rangle _{\phi _{i}}^{(A)}\langle
j|_{\phi _{i}}^{(A)}$, and the support of this function is contained in $%
D_{\phi _{i}}\cup F_{\phi _{i}}$. If $A_{j\phi _{i}}=0$, we can repeat the
argument with the operator $\widehat{A}_{\phi _{i}}+\alpha \widehat{B}_{\phi
_{i}}$ and take the limit $\alpha \rightarrow 0$, and we will find the same
result.

From eq.(\ref{PI.8}) we have 
\begin{equation}
\widehat{A}=\sum_{ij}A_{j\phi _{i}}|j\rangle _{\phi _{i}}^{(A)}\langle
j|_{\phi _{i}}^{(A)}  \label{PI.9}
\end{equation}
all over ${\cal M}$. Moreover, from eq.(\ref{PI.8}) we also have 
\[
symb\widehat{A}_{\phi _{i}}=\sum_{j}A_{j\phi _{i}}symb|j\rangle _{\phi
_{i}}^{(A)}\langle j|_{\phi _{i}}^{(A)} 
\]
and, as we have just proved, 
\[
symb|j\rangle _{\phi _{i}}^{(A)}\langle j|_{\phi _{i}}^{(A)}(\phi )=0\text{
if }\phi \notin D_{\phi _{i}}\subset D_{\phi _{i}}\cup F_{\phi _{i}} 
\]
Then, since for $i\neq k$, $D_{\phi _{i}}$ $\cap $ $D_{\phi _{k}}=\emptyset $
(but $F_{\phi _{i}}\cap F_{\phi _{j}}\neq 0$), we have 
\[
|\langle j|_{\phi _{i}}^{(A)}|j^{\prime }\rangle _{\phi
_{k}}^{(A)}|^{2}=\langle j|_{\phi _{i}}^{(A)}|j^{\prime }\rangle _{\phi
_{k}}^{(A)}\langle j^{\prime }|_{\phi _{k}}^{(A)}|j\rangle _{\phi
_{i}}^{(A)}=(|j^{\prime }\rangle _{\phi _{k}}^{(A)}\langle j^{\prime
}|_{\phi _{k}}^{(A)}||j\rangle _{\phi _{i}}^{(A)}\langle j|_{\phi
_{i}}^{(A)})= 
\]
\[
\int_{{\cal M}}symb|j^{\prime }\rangle _{\phi _{k}}^{(A)}\langle j^{\prime
}|_{\phi _{k}}^{(A)}symb|j\rangle _{\phi _{i}}^{(A)}\langle j|_{\phi
_{i}}^{(A)}d\phi ^{2(N+1)}= 
\]
\begin{equation}
\int_{F}symb|j^{\prime }\rangle _{\phi _{k}}^{(A)}\langle j^{\prime }|_{\phi
_{k}}^{(A)}symb|j\rangle _{\phi _{i}}^{(A)}\langle j|_{\phi _{i}}^{(A)}d\phi
^{2(N+1)}=0(\varepsilon ^{2(N+1)})  \label{33°}
\end{equation}
where $F$ is the union of all the joining zones $F_{\phi _{i}}$ and $%
\varepsilon $ is the characteristic width of the joining zone. Therefore,
for $i\neq k$ and $\varepsilon \rightarrow 0$,\footnote{%
Precisely: let us call $V_{{\cal M}}$ the volume of phase space: $V_{{\cal M}%
}\sim S^{N+1}$. Analogously, 
\[
I_{{\cal M}}=\int_{{\cal M}}symb|j^{\prime }\rangle _{\phi
_{k}}^{(A)}\langle j^{\prime }|_{\phi _{k}}^{(A)}symb|j\rangle _{\phi
_{i}}^{(A)}\langle j|_{\phi _{i}}^{(A)}d\phi ^{2(N+1)}\sim V_{{\cal M}}\sim
S^{N+1} 
\]
Let us also define $I_{\varepsilon }=\int_{{\cal F}}symb|j^{\prime }\rangle
_{\phi _{k}}^{(A)}\langle j^{\prime }|_{\phi _{k}}^{(A)}symb|j\rangle _{\phi
_{i}}^{(A)}\langle j|_{\phi _{i}}^{(A)}$ $d\phi ^{2(N+1)}\sim V_{\varepsilon
}=\varepsilon ^{2(N+1)}$. In order to prove eq.(\ref{33°}), it is necessary
that $I_{\varepsilon }\ll I_{{\cal M}}$ in such a way that $I_{\varepsilon }$
could be neglected. But $I_{\varepsilon }\sim V_{\varepsilon }$ and $I_{%
{\cal M}}\sim S^{N+1}$; then, $\varepsilon ^{2}\ll S$ $.$%
\par
Therefore, $\varepsilon $ must be:
\par
1.- Such that the ratio $\frac{\hbar }{\varepsilon ^{2}}$ be negligible to
eliminate the unwanted terms $\frac{i\hbar }{2}\overleftarrow{\partial }%
_{a}\omega ^{ab}\overrightarrow{\partial }_{b}$ in the joining zone (see
footnote 10).
\par
2.- As small as $\varepsilon ^{2}\ll S$ to satisfy eq.(\ref{33°}).
\par
Since $\hbar \ll S$, we can satisfy both conditions with an adequate $%
\varepsilon $, namely, such that: 
\[
\hbar \ll \varepsilon ^{2}\ll S 
\]
} we obtain 
\begin{equation}
\langle j|_{\phi _{i}}^{(A)}|j^{\prime }\rangle _{\phi _{k}}^{(A)}=0
\label{33'}
\end{equation}
This means that, in the limit $\varepsilon \rightarrow 0$, decomposition (%
\ref{PI.9}) is an {\it orthogonal decomposition} in the $|j\rangle _{\phi
_{i}}^{(A)}$.

{\bf Remark. }Let us now discuss the subspaces that can be defined by the
above decomposition. Being \{$|j\rangle _{\phi _{i}}^{(A)}\}$ a basis of the
Hilbert space ${\cal H}$ where we are working, any $|\varphi \rangle \in 
{\cal H}$ could be decomposed as 
\[
|\varphi \rangle =\sum_{ij}\varphi _{ij}|j\rangle _{\phi
_{i}}^{(A)}=\sum_{i}|\varphi _{i}\rangle 
\]
where 
\[
|\varphi _{i}\rangle =\sum_{j}\varphi _{ij}|j\rangle _{\phi _{i}}^{(A)} 
\]
and this ket belongs to subspace ${\cal H}_{i}$ of ${\cal H}$. Then, 
\begin{equation}
{\cal H=}\bigoplus_{i}{\cal H}_{i}  \label{I}
\end{equation}
Nevertheless, from eq.(\ref{PI.8}) where the off-diagonal terms $i\neq j$
are absent, we have 
\[
\widehat{A}_{\phi _{i}}\in {\cal H}_{i}\otimes {\cal H}_{i}={\cal O}_{i} 
\]
and from $\widehat{A}\in {\cal H}\otimes {\cal H}$ and eq.(\ref{PI.6}) we
have\footnote{%
This decomposition is similar to the decomposition of a function in its even
and odd parts in a Fourier transformation, where the $\sin $ is the even
basis and the $\cos $ is the odd basis.} 
\begin{equation}
{\cal O=H}\otimes {\cal H=}\bigoplus_{i}{\cal O}_{i}=\bigoplus_{i}{\cal H}%
_{i}\otimes {\cal H}_{i}  \label{II}
\end{equation}
This shows that {\it there are no cross terms} ${\cal H}_{i}\otimes {\cal H}%
_{j}$ in the decomposition of ${\cal O=H}\otimes {\cal H}$.

We can see that the decomposition that really matters for our discussion is (%
\ref{II}), the decomposition in subspaces ${\cal O}_{i}$, and not (\ref{I}):
the repeated index $i$ in the basis $\{|j\rangle _{\phi _{i}}^{(A)}\langle
j^{\prime }|_{\phi _{i}}^{(A)}\}$ means that this basis corresponds to the
decomposition done in the ${\cal O}_{i}$, which is the relevant one for this
paper.

\subsection{Decoherence in the energy}

a.- We will now introduce decoherence according to the self-induced
approach. Let us define, {\it in each }$D_{\phi _{i}}$, a local $N+1-$CSCO
where, as in eq.(\ref{PI.9}), the observables of the $N+1-$CSCO $\{\widehat{H%
},\widehat{O_{\phi _{i}}}\}$ are decomposed as 
\begin{equation}
\widehat{H}=\int_{0}^{\infty }\omega \sum_{im}|\omega ,m\rangle _{\phi
_{i}}\langle \omega ,m|_{\phi _{i}}d\omega ,\quad \text{ }\widehat{O_{\phi
_{i}I}}=\int_{0}^{\infty }\sum_{m}O_{m_{I\phi _{i}}}|\omega ,m\rangle _{\phi
_{i}}\langle \omega ,m|_{\phi _{i}}d\omega  \label{1.6}
\end{equation}
where the energy spectrum is $0\leq \omega <\infty $ and $m_{I\phi
_{i}}=\{m_{1\phi _{i}},...,m_{N\phi _{i}}\}$, $m_{I\phi _{i}}\in {\Bbb N}$
(the spectra of the $\widehat{O_{\phi _{i}I}}$ are discrete for simplicity).%
\footnote{%
Hamiltonians with continuous spectra are considered in papers \cite
{Laura-Cast 1998-E} and \cite{Laura-Cast 1998-A}. We use this kind of
spectra since they are the usual ones in the macroscopic limit $\hbar
\rightarrow 0$ (see \cite{Stockmann} eq.(3.1.24) p.67). Strictly, we should
call $|\omega ,m\rangle _{\phi _{i}}$ $^{(\widehat{H},\widehat{O_{\phi _{i}}}%
)}$ the vectors $|\omega ,m\rangle _{\phi _{i}}$, but we will just call them 
$|\omega ,m\rangle _{\phi _{i}}$ for simplicity.} Therefore 
\begin{equation}
\quad \widehat{H}|\omega ,m\rangle _{\phi _{i}}=\omega |\omega ,m\rangle
_{\phi _{i}},\quad \text{ }\widehat{O_{\phi _{i}I}}|\omega ,m\rangle _{\phi
_{i}}=O_{m_{I\phi _{i}}}|\omega ,m\rangle _{\phi _{i}}\quad  \label{1.7}
\end{equation}
where the $|\omega ,m\rangle _{\phi _{i}}$ are the eigenvectors of the
observables $\widehat{H},$ and $\widehat{O_{\phi _{i}}}$ (such that $symb%
\widehat{O_{\phi _{i}}}=O_{\phi _{i}}(\phi )\neq 0$ only in $D_{\phi
_{i}}\cap F_{\phi _{i}})$ and $m$ is a shorthand for $m_{\phi
_{i}I}=\{m_{\phi _{i}1},...,m_{\phi _{i}N}\}$. The set $\{|\omega ,m\rangle
_{\phi _{i}}\}$ is orthonormal in $\omega $ and in $m$, in the usual
eigenvalue indices and in $i$, as proved in eq.(\ref{33'}): 
\begin{equation}
\langle \omega ,m|_{\phi _{i}}|\omega ^{\prime },m^{\prime }\rangle _{\phi
_{j}}=\delta (\omega -\omega ^{\prime })\delta _{mm^{\prime }}\delta _{ij}
\label{1.7A}
\end{equation}

b.- Now we can define our relevant algebra of observables. This choice {\it %
will play the role of coarse-graining} in our approach. A generic observable
reads, in the orthonormal basis just defined, 
\begin{equation}
\widehat{O}=\sum_{imm^{\prime }}\int_{0}^{\infty }\int_{0}^{\infty }d\omega
d\omega ^{\prime }\widetilde{O}(\omega ,\omega ^{\prime })_{\phi
_{i}mm^{\prime }}|\omega ,m\rangle _{\phi _{i}}\langle \omega ^{\prime
},m^{\prime }|_{\phi _{i}}  \label{A1.1}
\end{equation}
where $\widetilde{O}(\omega ,\omega ^{\prime })_{\phi _{i}mm^{\prime }}$ is
a generic kernel or distribution in $\omega ,$ $\omega ^{\prime }$.\footnote{%
As explained at the end of the last subsection, the index $i$ in projector $%
|\omega ,m\rangle _{\phi _{i}}\langle \omega ^{\prime },m^{\prime }|_{\phi
_{i}}$ corresponds to the fact that the decomposition is done in the ${\cal O%
}_{i}$ and, therefore, the index is repeated in $|\omega ,m\rangle _{\phi
_{i}}$ and in $\langle \omega ^{\prime },m^{\prime }|_{\phi _{i}}$.} But we
must restrict this set of observables since it is too large for our
purposes; furthermore, it is not easy to work with generic kernels or
distributions. However, we cannot make the algebra too small either. In
fact, let us suppose that, in order to make computation easier, we postulate
that the $\widetilde{O}(\omega ,\omega ^{\prime })_{\phi _{i}mm^{\prime }}$
be just regular functions. Then, the states read 
\[
\widehat{\rho }=\sum_{imm^{\prime }}\int_{0}^{\infty }\int_{0}^{\infty
}d\omega d\omega ^{\prime }\overline{\widetilde{\rho }(\omega ,\omega
^{\prime })_{\phi _{i}mm^{\prime }}}|\omega ,m\rangle _{\phi }\langle \omega
^{\prime },m^{\prime }|_{\phi _{i}} 
\]
where the $\widetilde{\rho }(\omega ,\omega ^{\prime })_{\phi _{i}mm^{\prime
}},$ in the dual space, are also regular functions. Then, 
\[
\langle \widehat{O}\rangle _{\widehat{\rho }(t)}=\sum_{imm^{\prime
}}\int_{0}^{\infty }\int_{0}^{\infty }d\omega d\omega ^{\prime }\overline{%
\widetilde{\rho }(\omega ,\omega ^{\prime })_{\phi _{i}mm^{\prime }}}%
e^{i(\omega -\omega ^{\prime })t}\widetilde{O}(\omega ,\omega ^{\prime
})_{\phi _{i}mm^{\prime }} 
\]
and, since the product $\overline{\widetilde{\rho }(\omega ,\omega ^{\prime
})_{\phi _{i}mm^{\prime }}}\widetilde{O}(\omega ,\omega ^{\prime })_{\phi
_{i}mm^{\prime }}$ is a regular function (i.e. ${\Bbb L}_{1}$ in $\nu
=\omega -\omega ^{\prime })$, as a result of the Riemann-Lebesgue theorem
the mean value $\langle \widehat{O}\rangle _{\widehat{\rho }(t)}$ would
vanish for $t\rightarrow \infty $: we would obtain destructive interference
not only for the off-diagonal terms, {\it but for all of them}. On the
contrary, if $\overline{\widetilde{\rho }(\omega ,\omega ^{\prime })_{\phi
_{i}mm^{\prime }}}$ and $\widetilde{O}(\omega ,\omega ^{\prime })_{\phi
_{i}mm^{\prime }}$ were generic kernels, we could not use the
Riemann-Lebesgue theorem, and we can presume that there will be no
destructive interference. This means that $\widetilde{O}(\omega ,\omega
^{\prime })_{\phi _{i}mm^{\prime }}$ cannot be {\it so regular }nor{\it \ so
non-regular:} we must choose something in between. In order to avoid these
unacceptable results, the simplest choice is the van Hove choice; so, as in
paper \cite{Cast-Laura 2000-PRA}, we will take: 
\begin{equation}
\widetilde{O}(\omega ,\omega ^{\prime })_{\phi _{i}mm^{\prime }}=O(\omega
)_{\phi _{i}mm^{\prime }}\delta (\omega -\omega ^{\prime })+O(\omega ,\omega
^{\prime })_{\phi _{i}mm^{\prime }}  \label{A1.2}
\end{equation}
where the $O(\omega ,\omega ^{\prime })_{\phi _{i}mm^{\prime }}$ are
ordinary functions of the real variables $\omega $ and $\omega ^{\prime }$
(these functions must have some mathematical properties in order to develop
the theory; these properties are listed in paper \cite{Laura-Cast 1998-A}).
This choice is theoretically explained in papers \cite{van Hove 1955}, \cite
{van Hove 1956}, \cite{van Hove 1957}, \cite{van Hove 1959}, \cite{van Hove
1979}, \cite{Antoniou}, and \cite{Cast-Laura 2000-PRA}. Moreover, we need
the $\delta (\omega -\omega ^{\prime })$ term in order that the members of
the $N+1-$CSCO of eq.(\ref{1.6}) be contained in the space of observables.
So our operator belongs to an algebra $\widehat{{\cal A}}$ (defined by eq.(%
\ref{A1.2}) and the properties just required for the $O(\omega ,\omega
^{\prime })_{\phi _{i}mm^{\prime }}$), and reads 
\begin{equation}
\widehat{O}=\sum_{imm^{\prime }}\int_{0}^{\infty }d\omega O(\omega )_{\phi
_{i}mm^{\prime }}|\omega ,m\rangle _{\phi _{i}}\langle \omega ,m^{\prime
}|_{\phi _{i}}+\sum_{imm^{\prime }}\int_{0}^{\infty }\int_{0}^{\infty
}d\omega d\omega ^{\prime }O(\omega ,\omega ^{\prime })_{\phi _{i}mm^{\prime
}}|\omega ,m\rangle _{\phi _{i}}\langle \omega ^{\prime },m^{\prime }|_{\phi
_{i}}  \label{1.7'}
\end{equation}
The first term in the r.h.s. will be called $\widehat{O_{S}}$, the {\it %
singular} component, and the second term will be called $\widehat{O_{R}}$,
the {\it regular} component,\footnote{%
The component $\widehat{O}_{S}$ is called singular because it contains a
hidden distribution $\delta (\omega -\omega ^{\prime })$. In fact, it can be
obtained from the regular part by making $O(\omega ,\omega ^{\prime })_{\phi
_{i}mm^{\prime }}=O(\omega )_{\phi _{i}mm^{\prime }}\delta (\omega -\omega
^{\prime })$.} and $[\widehat{H},\widehat{O_{S}}]=0.$ The {\it observables}
are the self-adjoint $O^{\dagger }=O$ operators. We will say that these
observables belong to a space $\widehat{{\cal O}}$ (which is contained in
the operator algebra $\widehat{{\cal A}\text{ }}$ ); $\{|\omega ,m,m^{\prime
})_{\phi _{i}}$, $|\omega ,\omega ^{\prime },m,m^{\prime })_{\phi _{i}}\}$
is a basis of this space, where 
\begin{equation}
|\omega ,m,m^{\prime })_{\phi _{i}}\doteq |\omega ,m\rangle _{\phi
_{i}}\langle \omega ,m^{\prime }|_{\phi _{i}},\text{ \qquad }|\omega ,\omega
^{\prime },m,m^{\prime })_{\phi _{i}}\doteq |\omega ,m\rangle _{\phi
_{i}}\langle \omega ^{\prime },m^{\prime }|_{\phi _{i}}  \label{2.5'}
\end{equation}
Then, the classical analogue of eq.(\ref{1.7'}) would be 
\[
O(\phi )=\sum_{imm^{\prime }}\int_{0}^{\infty }d\omega O(\omega )_{\phi
_{i}mm^{\prime }}|\omega ,m,m^{\prime }(\phi ))_{\phi
_{i}}+\sum_{imm^{\prime }}\int_{0}^{\infty }\int_{0}^{\infty }d\omega
d\omega ^{\prime }O(\omega ,\omega ^{\prime })_{\phi _{i}mm^{\prime
}}|\omega ,\omega ^{\prime },m,m^{\prime }(\phi ))_{\phi _{i}} 
\]
where $|\omega ,m,m^{\prime }(\phi ))_{\phi _{i}}=symb|\omega ,m,m^{\prime
})_{\phi _{i}}$ and $|\omega ,\omega ^{\prime },m,m^{\prime }(\phi ))_{\phi
_{i}}=symb|\omega ,\omega ^{\prime },m,m^{\prime })_{\phi _{i}}$.

c.- The quantum states $\widehat{\rho }$ are measured by the observables
just defined, leading to the mean values of these observables; in the usual
notation: $\langle \widehat{O}\rangle _{\widehat{\rho }}=Tr(\widehat{\rho }%
^{\dagger }\widehat{O})$. We can conceive that mean values as the more
primitive objects of the quantum theory (see \cite{Ballentine}). These mean
values, generalized as in paper \cite{Laura-Cast 1998-A} and symbolized as $(%
\widehat{\rho }|\widehat{O})$, can be considered as the result of the action
of the linear functionals $\widehat{\rho }$ on the observables of the vector
space $\widehat{{\cal O}}$. Then, $\widehat{\rho }\in \widehat{{\cal S}}%
{\cal \subset }\widehat{{\cal O}}^{^{\prime }}$, where $\widehat{{\cal S}}$
is a convenient (i.e. satisfying eqs.(\ref{2.6º}) and (\ref{2.6'}) below)
convex set contained in $\widehat{{\cal O}}^{^{\prime }}$, the space of
linear functionals over $\widehat{{\cal O}}$. The basis of $\widehat{{\cal O}%
}^{\prime }$ (that is, the {\it co-basis} of $\widehat{{\cal O}}$ in each $%
D_{\phi _{i}}{\cal )}$ is \{$(\omega ,mm^{\prime }|_{\phi _{i}}$, $(\omega
\omega ^{\prime },mm^{\prime }|_{\phi _{i}}\}$, and it is defined in terms
of its functionals by the equations 
\[
\quad (\omega ,m,m^{\prime }|_{\phi _{i}}|\eta ,n,n^{\prime })_{\phi
_{j}}=\delta (\omega -\eta )\delta _{mn}\delta _{m^{\prime }n^{\prime
}}\delta _{ij} 
\]
\begin{equation}
(\omega ,\omega ^{\prime },m,m^{\prime }|_{\phi _{i}}|\eta ,\eta ^{\prime
},n,n^{\prime })_{\phi _{j}}=\delta (\omega -\eta )\delta (\omega ^{\prime
}-\eta ^{\prime })\delta _{mn}\delta _{m^{\prime }n^{\prime }}\delta _{ij}
\label{2.5''}
\end{equation}
and all other $(.|.)$ are zero. The orthogonality in $i,j,...$ is a
consequence of eqs.(\ref{1.7A}) and (\ref{2.5'}). Let us observe that $%
(\omega ,\omega ^{\prime },m,m^{\prime }|_{\phi _{i}}\doteq |\omega
,m\rangle _{\phi _{i}}\langle \omega ^{\prime },m^{\prime }|_{\phi _{i}}$
but $(\omega ,m,m^{\prime }|_{\phi _{i}}\neq |\omega ,m\rangle _{\phi
_{i}}\langle \omega ,m^{\prime }|_{\phi _{i}}$.\footnote{%
If $(\omega ,m,m^{\prime }|_{\phi _{i}}=|\omega ,m\rangle _{\phi
_{i}}\langle \omega ,m^{\prime }|_{\phi _{i}}$, it is easy to show that a
divergence appears.} Then, a generic quantum state reads 
\begin{eqnarray}
&&  \nonumber \\
\widehat{\rho }=\sum_{imm^{\prime }}\int_{0}^{\infty }d\omega \overline{\rho
(\omega )}_{\phi _{i}mm^{\prime }}(\omega ,mm^{\prime }|_{\phi _{i}}
&&+\sum_{imm^{\prime }}\int_{0}^{\infty }d\omega \int_{0}^{\infty }d\omega
^{\prime }\overline{\rho (\omega ,\omega ^{\prime })}_{\phi _{i}mm^{\prime
}}(\omega \omega ^{\prime },mm^{\prime }|_{\phi _{i}}  \label{1.8}
\end{eqnarray}
This is also the case for the corresponding classical analogue $\rho (\phi )$
of $\widehat{\rho }$, 
\begin{eqnarray}
&&  \nonumber \\
&&\rho (\phi )=\sum_{imm^{\prime }}\int_{0}^{\infty }d\omega \overline{\rho
(\omega )}_{\phi _{i}mm^{\prime }}(\omega ,mm^{\prime }(\phi )|_{\phi
_{i}}+\sum_{imm^{\prime }}\int_{0}^{\infty }d\omega \int_{0}^{\infty
}d\omega ^{\prime }\overline{\rho (\omega ,\omega ^{\prime })}_{\phi
_{i}mm^{\prime }}(\omega \omega ^{\prime },mm^{\prime }(\phi )|_{\phi _{i}}
\label{1.8'}
\end{eqnarray}
Each of the terms of the sum $\sum_{i}$ can be considered as a term of a
decomposition, where each of the $\rho _{\phi _{i}}(\phi ,t)=symb\widehat{%
\rho }_{\phi _{i}}(t)$ does not vanish in the corresponding domain $D_{\phi
_{i}}\subset {\cal D}_{\phi _{i}}$.\footnote{%
Considering for a moment the larger domains ${\cal D}_{\phi _{i}}$, of
course there are other charts defined in other domains ${\cal D}_{\phi
_{k}}^{\prime }$ and, therefore, other $D_{\phi _{k}}^{\prime }\subset {\cal %
D}_{\phi _{k}}^{\prime }$. But since $\rho (\phi ,t)=symb\widehat{\rho }(t)$
is defined in the whole ${\cal M}$, the ${\cal D}_{\phi _{i}}$, ${\cal D}%
_{\phi _{k}}^{\prime }$ are just {\it local} charts for which the same
function $\rho (\phi ,t)=symb\widehat{\rho }(t)$ is {\it globally} defined
in phase space. Moreover, at $\phi \in $ ${\cal D}_{\phi _{i}}$ $\cap $ $%
{\cal D}_{\phi _{j}}$, any pair of charts can be $C^{r}-$smoothly connected
in the sense that all their elements can be smoothly connected among each
other. The same argument can be applied to the partial decompositions of $%
\rho _{R}(\phi ,t)=symb\widehat{\rho }_{R}(t),$ $\rho _{S}(\phi ,t)=symb%
\widehat{\rho }_{S}(t)$ (i.e. the first and second terms of the r.h.s. of
the last equation), since both regular and singular parts are also {\it %
globally} defined in phase space.} As before, the first term in the r.h.s.
of eq.(\ref{1.8}) will be called $\widehat{\rho _{S}}$, the {\it singular}
component, and the second term will be called $\widehat{\rho _{R}}$, the 
{\it regular} component. Functions $\rho (\omega ,\omega ^{\prime })_{\phi
_{i}mm^{\prime }}$ are regular (see \cite{Laura-Cast 1998-A} for details).

Going back to the decomposition (\ref{1.8'}), we impose the following
conditions. We require that $\widehat{\rho }^{\dagger }=\widehat{\rho }$,
i.e. 
\begin{equation}
\quad \overline{\rho (\omega ,\omega ^{\prime })}_{\phi _{i}mm^{\prime
}}=\rho (\omega ^{\prime },\omega )_{\phi _{i}m^{\prime }m}  \label{2.6º}
\end{equation}
and that $\overline{\rho (\omega )}_{\phi _{i}mm^{\prime }}$ be {\it real
and non-negative}, satisfying the total probability condition 
\begin{equation}
(\widehat{\rho }|\widehat{I})=\sum_{im}\int_{0}^{\infty }d\omega \rho
(\omega )_{\phi _{i}}=1  \label{2.6'}
\end{equation}
where $\widehat{I}=\int_{0}^{\infty }d\omega \sum_{im}|\omega ,m\rangle
_{\phi _{i}}\langle \omega ,m|_{\phi _{i}}$ is the identity operator (\ref
{PI.1}) in $\widehat{{\cal O}}$ represented in each $D_{\phi _{i}}$. Eq.(\ref
{2.6'}) is the extension to state functionals of the usual condition $Tr\rho
^{\dagger }=1$, when $\rho $ is a density operator. Thus, from now on, $%
Tr\rho \doteq (\rho |I)$. For these reasons, $\widehat{\rho }$ belongs to
the already defined convex set $\widehat{{\cal S}}\subset \widehat{O^{\prime
}}$. The time evolution of the quantum state $\widehat{\rho }$ reads 
\begin{equation}
\widehat{\rho }(t)=\sum_{imm^{\prime }}\int_{0}^{\infty }d\omega \overline{%
\rho (\omega )}_{\phi _{i}mm^{\prime }}(\omega ,mm^{\prime }|_{\phi
_{i}}+\sum_{imm^{\prime }}\int_{0}^{\infty }d\omega \int_{0}^{\infty
}d\omega ^{\prime }\overline{\rho (\omega ,\omega ^{\prime })}_{\phi
_{i}mm^{\prime }}e^{i(\omega -\omega ^{\prime })t/\hbar }(\omega \omega
^{\prime },mm^{\prime }|_{\phi _{i}}  \label{2.6''}
\end{equation}

As we have already said, at the statistical quantum level we can only
measure mean values of observables in quantum states 
\begin{equation}
\langle \widehat{O}\rangle _{\widehat{\rho }(t)}=(\widehat{\rho }(t)|%
\widehat{O})=\sum_{imm^{\prime }}\int_{0}^{\infty }d\omega \overline{\rho
(\omega )}_{\phi _{i}mm^{\prime }}O(\omega )_{\phi _{i}mm^{\prime
}}+\sum_{imm^{\prime }}\int_{0}^{\infty }d\omega \int_{0}^{\infty }d\omega
^{\prime }\overline{\rho (\omega ,\omega ^{\prime })}_{\phi _{i}mm^{\prime
}}e^{i(\omega -\omega ^{\prime })t/\hbar }O(\omega ,\omega ^{\prime })_{\phi
_{i}mm^{\prime }}  \label{1.9}
\end{equation}
From eq.(\ref{L.3.8}), the classical analogue has exactly the same form 
\begin{eqnarray}
\langle O(\phi )\rangle _{\rho (\phi ,t)} &=&(\rho (\phi ,t)|O(\phi
))=\sum_{imm^{\prime }}\int_{0}^{\infty }d\omega \overline{\rho (\omega )}%
_{\phi _{i}mm^{\prime }}O(\omega )_{\phi _{i}mm^{\prime }}+  \nonumber \\
&&+\sum_{imm^{\prime }}\int_{0}^{\infty }d\omega \int_{0}^{\infty }d\omega
^{\prime }\overline{\rho (\omega ,\omega ^{\prime })}_{\phi _{i}mm^{\prime
}}e^{i(\omega -\omega ^{\prime })t/\hbar }O(\omega ,\omega ^{\prime })_{\phi
_{i}mm^{\prime }}  \label{l.10}
\end{eqnarray}
Both decompositions are valid in each $D_{\phi _{i}}$. If we take into
account that $O(\omega ,\omega ^{\prime })$ and $\overline{\rho (\omega
,\omega ^{\prime })}_{\phi _{i}mm^{\prime }}$ are regular (as regular as
needed to use the Riemann-Lebesgue theorem, i. e. $O(\omega ,\omega ^{\prime
})$ $\overline{\rho (\omega ,\omega ^{\prime })}_{\phi _{i}mm^{\prime }}\in 
{\Bbb L}_{1}(\omega -\omega ^{\prime })$, see \cite{Laura-Cast 1998-A}), we
can take the limit $t\rightarrow \infty $ and use the Riemann-Lebesgue
theorem. As the result, we see that the fluctuating-regular part vanishes
and we arrive to the weak (quantum and classical) limits 
\[
W\lim_{t\rightarrow \infty }\widehat{\rho }(t)=\widehat{\rho _{S}}=\widehat{%
\rho }_{*}=\sum_{imm^{\prime }}\int_{0}^{\infty }d\omega \overline{\rho
(\omega ,p)}_{\phi _{i}mm^{\prime }}(\omega ,m,m^{\prime }|_{\phi _{i}} 
\]
\begin{equation}
W\lim_{t\rightarrow \infty }\rho (\phi ,t)=\rho _{S}(\phi )=\rho _{*}(\phi
)=\sum_{imm^{\prime }}\int_{0}^{\infty }d\omega \overline{\rho (\omega ,p)}%
_{\phi _{i}mm^{\prime }}(\omega ,m,m^{\prime },(\phi )|_{\phi _{i}}
\label{WLM}
\end{equation}
where $(\omega ,m,m^{\prime },(\phi )|_{\phi _{i}}=symb(\omega ,mm^{\prime
}|_{\phi _{i}}$ and $\rho _{*}(\phi )=symb\widehat{\rho }_{*}$ are defined
in ${\cal M}$ and the integral in eq.(\ref{WLM}) is decomposed in different
ways at each $D_{\phi _{i}}$. Since only the singular diagonal terms remain,
we have obtained decoherence in the energy variable $\omega .$ Precisely,
any quantum state weakly tends to a linear combination of the energy
diagonal states $(\omega ,m,m^{\prime }|_{\phi _{i}}$ (the energy
'off-diagonal' states $(\omega ,\omega ^{\prime },m,m^{\prime }|_{\phi _{i}}$
are not present in $\rho _{*}$). This is the case if we observe and measure
the system evolution with {\it any possible observable of space }$\widehat{%
{\cal O}}${\it .} Therefore, from the observational point of view, we have
decoherence of the energy levels in spite of the fact that, from the strong
limit point of view, the off-diagonal terms {\it never vanish:} they just
oscillate since we cannot directly use the Riemann-Lebesgue theorem in the
operator equation (\ref{2.6''}).

{\bf Important remarks}

i.- It may be supposed that decoherence takes place without a
coarse-graining. It is no so: the choice of the algebra $\widehat{{\cal A}}$
among all possible algebras (see under eq.(\ref{A1.2})) and the systematic
use of mean values $\langle \widehat{O}\rangle _{\widehat{\rho }(t)}=(%
\widehat{\rho }(t)|\widehat{O})$ (eq.(\ref{1.9})), restrict the available
information and produce the effect of a coarse-graining. In fact, we can
define the projector $\Pi =|\widehat{O})(\widehat{\rho }_{0}|$, with $|%
\widehat{O})\in \widehat{{\cal A}}$ and $(\widehat{\rho }_{0}|\widehat{O})=1$%
, that projects $(\widehat{\rho }(t)|$ as $(\widehat{\rho }(t)|\Pi =\langle 
\widehat{O}\rangle _{\widehat{\rho }(t)}(\widehat{\rho }_{0}|,$ and
translates everything in projectors language: we obtain, from eq.(\ref{WLM}%
), $\lim_{t\rightarrow \infty }(\widehat{\rho }(t)|\Pi =(\widehat{\rho }%
_{*}|\Pi $. This projection will obviously break the unitarity of the
primitive evolution. In this way we could develop a formalism closer to the
usual one. See a detailed explanation in \cite{SHPMP} and \cite
{Cast-Lombardi 2005-PS}.

ii.- Theoretically, decoherence takes place at $t\rightarrow \infty $. But,
in practice, decoherence appears at a decoherence time, as we have defined
in \cite{Rolo}: the decoherence time can be easily computed from the poles
of the resolvent or the initial conditions density in the complex extension
of the $\widehat{H}$ spectrum. Trivial $\widehat{H}$ (e.g. free particle $%
\widehat{H}$) and trivial initial conditions (e.g. zero temperature ones) do
not have poles and the decoherence time is infinite. This means that, to
reach equilibrium in a finite characteristic time, $\widehat{H}$ must be
non-trivial (e.g. the sum of a free Hamiltonian plus an interaction
Hamiltonian) and/or the initial conditions must be non-trivial (e.g. $T\neq
0 $). For details, see \cite{Cast-Lombardi 2005}, where decoherence times
are estimated in $10^{-13}-10^{-15}s$\ for microscopic bodies, and $%
10^{-37}-10^{-39}s$\ for macroscopic bodies; for a thermal bath our results
coincide with those obtained by the einselection approach.

\subsection{Decoherence in the remaining variables}

Having obtained decoherence in the energy levels, we must consider
decoherence in the other dynamical variables $O_{\phi _{i}I}$ of the set of
local CSCOs we are using. We will call these variables 'momentum variables'.
Since the expression of $\rho _{*}$, given in eq.(\ref{WLM}), only involves
the time independent components of $\rho (t)$, it is impossible that a
further decoherence process eliminates the off-diagonal terms in the
remaining $N$ dynamical momentum variables. Therefore, the only alternative
is to find the basis where these off-diagonal components $\rho (\omega
)_{\phi _{i}mm^{\prime }}$ vanish at any time.

Let us consider the following unitary change of basis 
\begin{equation}
\qquad |\omega ,p\rangle _{\phi _{i}}=\sum_{m}U(\omega )_{mp}|\omega
,m\rangle _{\phi _{i}}  \label{2.11'}
\end{equation}
where $p$ and $m$ are shorthand notations for $p\doteq \{p_{1},...,p_{N}\}$
and $m\doteq \{m_{1},...,m_{N}\}$, and $\left[ U(\omega )^{-1}\right] _{mp}=%
\overline{U(\omega )}_{pm}$. We choose the new basis $\{|\omega ,p\rangle
_{\phi _{i}}\}$ such that it verifies the generalized orthogonality
condition 
\[
\quad \langle \omega ,p|_{\phi _{i}}|\omega ^{\prime },p^{\prime }\rangle
_{\phi _{i}}=\delta (\omega -\omega ^{\prime })\delta _{pp^{\prime }} 
\]
Since $\overline{\rho (\omega )}_{\phi _{i}}=\rho (\omega )_{\phi _{i}}$, it
is possible to choose $U(\omega )$ in such a way that the off-diagonal parts
of $\rho (\omega )_{\phi _{i}pp^{\prime }}$ vanish, i.e. 
\begin{equation}
\qquad \rho (\omega )_{\phi _{i}pp^{\prime }}=\rho _{\phi _{i}}(\omega
)_{p}\,\delta _{pp^{\prime }}  \label{nonint}
\end{equation}
This means that there is a{\it \ final local pointer basis in }$D_{\phi
_{i}} $ for the observables, given by $\{|\omega ,p,p^{\prime })_{\phi _{i}}$%
, $|\omega ,\omega ^{\prime },p,p^{\prime })_{\phi _{i}}\}$ and defined as
in eq.(\ref{2.5'}) but now with the $p$. The corresponding final pointer
basis for the states, $\{(\omega ,p,p^{\prime }|_{\phi _{i}}$, $(\omega
,\omega ^{\prime },p,p^{\prime }|_{\phi _{i}}\}$, diagonalizes the time
independent part of $\rho (t)$ and, therefore, diagonalizes the final state $%
\rho _{*}$

Now, we have diagonalized the $\overline{\rho (\omega )}_{\phi
_{i}mm^{\prime }}$ in $m$ and $m^{\prime }$, obtaining 
\[
W\lim_{t\rightarrow \infty }\widehat{\rho }(t)=\widehat{\rho _{S}}=\widehat{%
\rho }_{*}=\sum_{ip}\int_{0}^{\infty }d\omega \overline{\rho _{\phi
_{i}}(\omega )}_{p}(\omega ,p,p|_{\phi _{i}} 
\]
\begin{equation}
W\lim_{t\rightarrow \infty }\rho (\phi ,t)=\rho _{S}(\phi )=\rho _{*}(\phi
)=\sum_{ip}\int_{0}^{\infty }d\omega \overline{\rho _{\phi _{i}}(\omega )}%
_{p}(\omega ,p,p,(\phi )|_{\phi _{i}}  \label{1.12}
\end{equation}
Here we are using a local pointer $N+1-$CSCO $\{\widehat{H},\widehat{P}%
_{\phi _{i}1},...,\widehat{P}_{\phi _{i}N}\}$ at each $D_{\phi _{i}}$, where
the $\widehat{P}_{\phi _{i}I}$ are

\begin{equation}
\widehat{P}_{\phi _{i}I}=\sum_{i}\int_{0}^{\infty }d\omega \sum_{p}p_{\phi
_{i}I}(\omega )|\omega ,p,p)_{\phi _{i}}  \label{1.13}
\end{equation}
and their classical analogues 
\begin{equation}
P(\phi )_{\phi _{i}I}=\sum_{i}\int_{0}^{\infty }d\omega \sum_{p}p_{\phi
_{i}I}(\omega )|\omega ,p,p(\phi ))_{\phi _{i}}  \label{l.13'}
\end{equation}
where $|\omega ,p,p)_{\phi _{i}}=|\omega ,p\rangle _{\phi _{i}}\langle
\omega ,p|_{\phi _{i}}$ or simply $\{|\omega ,p\rangle _{\phi _{i}}\}$ is
the {\it local pointer basis} in $D_{\phi _{i}}$; so, we can write eq.(\ref
{1.7'}) in this new basis (see eq.(\ref{L.2.2}) below).\footnote{%
The complexity of these formulae demonstrates why it was so difficult to
define the pointer basis in a general case. As we can see, the pointer basis
depends on $H$ and the initial conditions, but there are some cases (see
section IV) where it only depends on $H$.} Now all the operators and
matrices involved are diagonal, and decoherence is complete. We can define
all the observables $\widehat{O\text{ }}$of eq.(\ref{1.7'}) in this new
local pointer basis.

Since in the limit $\hbar \rightarrow 0$ we usually have $\widehat{P}$ with
continuous spectra, instead of the last equations we would have the natural
analogues of eqs.(\ref{1.12}) (see \cite{Cast-Gadella 2003} and \cite{Cast
2004} for details) 
\[
W\lim_{t\rightarrow \infty }\widehat{\rho }(t)=\widehat{\rho _{S}}=\widehat{%
\rho }_{*}=\sum_{i}\int_{0}^{\infty }d\omega \int_{p\epsilon D_{\phi
_{i}}}dp^{N}\overline{\rho (\omega )_{\phi _{i}}}(\omega ,p,p|_{\phi _{i}} 
\]
\begin{equation}
W\lim_{t\rightarrow \infty }\rho (\phi ,t)=\rho _{S}(\phi )=\rho _{*}(\phi
)=\sum_{i}\int_{0}^{\infty }d\omega \int_{p\epsilon D_{\phi _{i}}}dp^{N}%
\overline{\rho (\omega )_{\phi _{i}}}(\omega ,p,p,(\phi )|_{\phi _{i}}
\label{WLP}
\end{equation}
In the next section we will consider the classical limit and, then, we will
only use continuous spectra and equations like the last two:\footnote{%
If we use the Heisenberg picture, the $\widehat{A\text{ }}$ would become
diagonal. So, heuristically 
\[
\lim_{t\rightarrow \infty }(\widehat{\rho }_{*}|[\widehat{A}(t),\widehat{B}%
])=\lim_{t\rightarrow \infty }Tr(\widehat{\rho }_{*}\widehat{A}(t)\widehat{B}%
-\widehat{\rho }_{*}\widehat{B}\widehat{A}(t))= 
\]
\[
Tr(\widehat{\rho }_{*}\widehat{A}_{*}\widehat{B}-\widehat{\rho }_{*}\widehat{%
B}\widehat{A}_{*})=Tr(\widehat{\rho }_{*}\widehat{A}_{*}\widehat{B}-\widehat{%
A}_{*}\widehat{\rho }_{*}\widehat{B})=0 
\]
where $\widehat{A}_{*}$ is the diagonal weak limit of $\widehat{A}$ and,
therefore, commutes with the diagonal $\widehat{\rho }_{*}$. As a
consequence, the evolution is{\it \ (heuristically) weakly asymptotically
abelian }(\cite{Benatti}, Def. 4.11) since, in the limit $t\rightarrow
\infty $, $\widehat{{\cal A}}$ can be considered commutative. Therefore, a
quantum system with continuous spectrum is {\it weakly asymptotically abelian%
}.} so we will re-write some equations in the new basis for the sake of
completeness.

\section{The classical statistical limit}

\subsection{Quantum and classical operators}

a.- From now on, we will consider a system from the point of view of the
local pointer complete set of $N+1-$commuting observables $\{\widehat{H,}%
\widehat{P_{\phi _{i}1}},...,\widehat{P}_{\phi _{i}N}\}$, defined in eqs.(%
\ref{1.6}) and (\ref{1.13}). As above, to simplify the notation we will just
call $\{\widehat{H,}\widehat{P}_{\phi _{i}}\}$ the set $\{\widehat{H,}%
\widehat{P_{\phi _{i}1}},...,\widehat{P}_{\phi _{i}N}\}$. Thus, we will
consider the orthonormal eigenbasis $\{|\omega ,p\rangle _{\phi _{i}}\}$ of $%
\{\widehat{H,}\widehat{P}_{\phi _{i}}\}$,\footnote{%
We are using the `final pointer basis' of section III.C. Below we will write
all the formulae in this basis.} and write the Hamiltonian and $\widehat{P}$
as 
\begin{equation}
\widehat{H\text{ }}=\sum_{i}\int_{p\epsilon D_{\phi
_{i}}}dp^{N}\int_{0}^{\infty }\omega |\omega ,p\rangle _{\phi _{i}}\langle
\omega ,p|_{\phi _{i}}d\omega \qquad \widehat{P}_{\phi _{i}}=\int_{p\epsilon
D_{\phi _{i}}}dp^{N}\int_{0}^{\infty }p|\omega ,p\rangle _{\phi _{i}}\langle
\omega ,p|_{\phi _{i}}d\omega  \label{L.2.1}
\end{equation}
Furthermore, we will consider the algebra $\widehat{{\cal A}}$ of the
operators (\ref{1.7'}), which now read 
\[
\widehat{O}=\sum_{i}\int_{p\epsilon D_{\phi _{i}}}dp^{N}\int_{0}^{\infty
}O_{\phi _{i}}(\omega ,p)|\omega ,p\rangle _{\phi _{i}}\langle \omega
,p|_{\phi _{i}}d\omega 
\]
\begin{equation}
+\sum_{i}\int_{p\epsilon D_{\phi _{i}}}\int_{p\epsilon D_{\phi
_{i}}}dp^{N}dp^{\prime N}\int_{0}^{\infty }\int_{0}^{\infty }O_{\phi
_{i}}(\omega ,\omega ^{\prime },p,p^{\prime })|\omega ,p\rangle _{\phi
_{i}}\langle \omega ,p^{\prime }|_{\phi _{i}}d\omega d\omega ^{\prime }
\label{L.2.2}
\end{equation}
As before, the first term in the r.h.s. will be called $\widehat{O_{S}}$ ,
the {\it singular} component, and the second term will be called $\widehat{%
O_{R}}$, the {\it regular} component. Also as before, functions $O_{\phi
_{i}}(\omega ,\omega ^{\prime },p,p^{\prime })$ are regular (see \cite
{Laura-Cast 1998-A} for details), $[\widehat{H},\widehat{O_{S}}]=0$, $%
\widehat{O_{S}}\in \widehat{{\cal L}}_{S}$, where $\widehat{{\cal L}}_{S}$
is the singular space, $\widehat{O_{R}}\in \widehat{{\cal L}}_{R}$, where $%
\widehat{{\cal L}}_{R}$ is the regular space, and $\widehat{{\cal A}}=%
\widehat{{\cal L}}_{S}\oplus \widehat{{\cal L}}_{R}$. The observables are
the self-adjoint operators of $\widehat{{\cal A}}$, and they belong to a
space $\widehat{{\cal O}}.$

b.- Let us now consider the Wigner transformation of these objects. The
operators of $\widehat{{\cal L}}_{R}$ are regular; so, their transformation
is obtained as explained in Section II. Then, we have to consider only the
singular space $\widehat{{\cal L}}_{S}$, the space of the operators that
commute with $\widehat{H}$. This is not a regular space of operators on a
Hilbert space ${\cal H}$ as $\widehat{{\cal L}}_{R}$, since it contains a
hidden $\delta (\omega -\omega ^{\prime }$ ) (see eq.(\ref{A1.2})), but the
mapping $symb$ given by eq.(\ref{L.2.6}) can also be well defined for the
observables in $\widehat{{\cal L}}_{S}$. In fact, from eq.(\ref{L.2.2}) we
know that 
\begin{equation}
\widehat{O}_{S}=\sum_{i}\int_{p\epsilon D_{\phi _{i}}}dp^{N}\int_{0}^{\infty
}O_{\phi _{i}}(\omega ,p)|\omega ,p\rangle _{\phi _{i}}\langle \omega
,p|_{\phi _{i}}d\omega  \label{L.2.15}
\end{equation}
If we consider, as usual, first $O_{\phi _{i}}$ as a polynomial, and then $%
O_{\phi _{i}}$ as a function of a certain space where the polynomials are
dense,\footnote{%
These polynomials have several variables, but there is no problem since all
these variables commute.} by using eqs.(\ref{L.2.1}) we can conclude that 
\begin{equation}
\widehat{O}_{S}=\sum_{i}O_{\phi _{i}}(\widehat{H},\widehat{P_{\phi _{i}})}%
=\sum_{i}\widehat{O}_{S\phi _{i}}  \label{L.2.18}
\end{equation}
But, when $\widehat{f},\widehat{g}$ commute, as the members of the CSCO do
(see eq.(\ref{L.2.14})), 
\begin{equation}
symb(\widehat{f},\widehat{g})=(f*g)(\phi )=f(\phi )g(\phi )+0(\hbar ^{2})
\label{L.2.19}
\end{equation}
Then, by means of the same procedure as before and eq.(\ref{L.2.9})), 
\begin{equation}
symb\widehat{O}_{S}=O_{S}(\phi )=\sum_{i}O_{\phi _{i}}(H(\phi ),P_{\phi
_{i}}(\phi ))+0(\hbar ^{2})=\sum_{i}symb\widehat{O}_{S\phi _{i}}
\label{L.2.20}
\end{equation}
where $H(\phi ),P_{\phi _{i}}(\phi )$ can be computed as usually (see \cite
{Cast-Gadella 2003} for details). In this way, we have succeeded in
computing all the $symb$ of the observables of $\widehat{{\cal L}}_{S}$ up
to $0(\hbar ^{2})$, which are just the $O_{\phi _{i}}(H(\phi ),P(\phi ))$,
and we have defined the mapping

\begin{equation}
symb:\widehat{{\cal L}}_{S}\rightarrow {\cal L}_{Sq}\qquad symb\widehat{O_{S}%
}=O_{S}(\phi )=\sum_{i}O_{\phi _{i}}(H(\phi ),P_{\phi _{i}}(\phi ))+0(\hbar
^{2})  \label{L.2.22}
\end{equation}
Moreover, since decompositions $D_{\phi _{i}}$ or ${\cal D}_{\phi _{i}}$ are
arbitrary (because they depend on the initial conditions of Section III.A),
from eqs.(\ref{L.2.18}) and (\ref{L.2.20}) we obtain (up to $0(\hbar ^{2})$) 
\begin{equation}
\text{ }\widehat{O}_{S\phi _{i}}=O_{\phi _{i}}(\widehat{H},\widehat{P}_{\phi
_{i}}),\quad O_{S\phi _{i}}(\phi )=symb\widehat{O}_{S\phi _{i}}=O_{\phi
_{i}}(H(\phi ),P_{\phi _{i}}(\phi ))  \label{interp}
\end{equation}
Let us observe that, if $O_{\phi _{i}}(\omega ,p)=\delta (\omega -\omega
^{\prime })\delta (p-p^{\prime })$, we have (also up to $0(\hbar ^{2})$) 
\begin{equation}
symb|\omega ^{\prime },p^{\prime }\rangle _{\phi _{i}}\langle \omega
^{\prime },p^{\prime }|_{\phi _{i}}=\delta (H(\phi )-\omega ^{\prime
})\delta (P_{\phi _{i}}(\phi )-p)  \label{L.2.21}
\end{equation}
an equation that we will use below.

Summing up, from eqs.(\ref{L.2.4}) and (\ref{L.2.22}) we have defined a
classical space ${\cal A}_{q}{\cal =L}_{R}\oplus {\cal L}_{S}$ and a mapping

\begin{equation}
symb:\widehat{{\cal A}}\rightarrow {\cal A}_{q}\qquad symb\widehat{O}=O(\phi
)  \label{L.2.23}
\end{equation}
where eqs.(\ref{L.2.11}) and (\ref{L.2.12}) are also valid. Then, we can
repeat what we have said below eq.(\ref{L.2.12}), but now for the algebra $%
\widehat{{\cal A}}_{q}$ defined as in this section, with its {\it regular
and singular parts.}

If now we take the limit $\hbar \rightarrow 0$, we obtain ${\cal A}%
_{q}\rightarrow {\cal A}$, where ${\cal A}$ is the usual algebra of
observables on phase space. Then, in this limit we have a correspondence $%
\widehat{{\cal A}}\rightarrow {\cal A}$. However, even if this limit is well
defined and can be considered as {\it the classical limit of the algebra of
operators,} it is only the limit of the {\it equations} of the system, since
these are a consequence of the algebra. Therefore, this is just a 'formal'
limit. The limit $\hbar \rightarrow 0$ will be completely studied when we
deal with the state space.

For the sake of simplicity, from now on we will systematically eliminate all
the $0(\hbar ^{2})$ from the equations and call the ${\cal A}_{q}$ just $%
{\cal A}$. This is a rigorous simplification. In fact, when $\hbar =0$ we
can make the $0(\hbar )=0$ everywhere since, from eq.(\ref{L.2.9}), when $%
\hbar =0$ we have $\exp 0=1$ in that equation; in other words, the $%
\lim_{\hbar \rightarrow 0}$ is continuous.

\subsection{Quantum and classical states}

a.- Let us remember that $|\omega ,p)_{\phi _{i}}=|\omega ,p,p)_{\phi
_{i}}=|\omega ,p\rangle _{\phi _{i}}\langle \omega ,p|_{\phi _{i}}$ and $%
|\omega ,\omega ^{\prime },p,p^{\prime })_{\phi _{i}}=|\omega ,p\rangle
_{\phi _{i}}\langle \omega ^{\prime },p^{\prime }|_{\phi _{i}}$ as in eq.(%
\ref{2.5'}). $\{|\omega ,p,p^{\prime })_{\phi _{i}}\}$ is the basis of $%
\widehat{{\cal L}}_{S}$ and $\{|\omega ,\omega ^{\prime },p,p^{\prime
})_{\phi _{i}}\}$ is the basis of $\widehat{{\cal L}}_{R}$. Then, eq.(\ref
{L.2.2}) reads 
\[
\widehat{O}=\sum_{i}\widehat{O}=\sum_{i}\int_{p\epsilon D_{\phi
_{i}}}dp^{N}\int_{0}^{\infty }O_{\phi _{i}}(\omega ,p)|\omega ,p)_{\phi
_{i}}d\omega 
\]

\begin{equation}
+\sum_{i}\int_{p\epsilon D_{\phi _{i}}}\int_{p\epsilon D_{\phi
_{i}}}dp^{N}dp^{\prime N}\int_{0}^{\infty }\int_{0}^{\infty }O_{\phi
_{i}}(\omega ,\omega ^{\prime },p,p^{\prime })|\omega ,\omega ^{\prime
},p,p^{\prime })_{\phi _{i}}d\omega d\omega ^{\prime }  \label{L.3.1}
\end{equation}
Since the states are functionals over the space $\widehat{{\cal A}}=\widehat{%
{\cal L}}_{S}\oplus \widehat{{\cal L}}_{R}$, let us consider the dual space $%
\widehat{{\cal A}^{\prime }}=\widehat{{\cal L}_{S}^{\prime }}\oplus \widehat{%
{\cal L}^{\prime }}_{R}$. We will call $\{(\omega ,p|_{\phi _{i}}\}$ the
local bases of $\widehat{{\cal L}_{S}^{\prime }}$ and $\{(\omega ,\omega
^{\prime },p,p^{\prime }|_{\phi _{i}}\}$ the local bases of $\widehat{{\cal L%
}_{R}^{\prime }}.$ Let us remember that $(\omega ,\omega ^{\prime
},p,p^{\prime }|_{\phi _{i}}=|\omega ,p\rangle _{\phi _{i}}\langle \omega
^{\prime },p^{\prime }|_{\phi _{i}}$ but $(\omega ,p|_{\phi _{i}}\neq
|\omega ,p\rangle _{\phi _{i}}\langle \omega ,p|_{\phi _{i}}$. Moreover, as
in eq.(\ref{2.5''}), 
\[
(\omega ,p|_{\phi _{i}}|\omega ^{\prime },p^{\prime })_{\phi j}=\delta
(\omega -\omega ^{\prime })\delta ^{N}(p-p^{\prime })\delta _{ij}\qquad
(\omega ,\sigma ,p,s|_{\phi _{i}}|\omega ^{\prime },\sigma ^{\prime
},p^{\prime },s^{\prime })_{\phi _{j}}=\delta (\omega -\omega ^{\prime
})\delta (\sigma -\sigma ^{\prime })\delta ^{N}(p-p^{\prime })\delta
^{N}(s-s^{\prime })\delta _{ij} 
\]
\begin{equation}
(\omega ,\sigma ,|_{\phi _{i}}|\omega ^{\prime },\sigma ^{\prime },p^{\prime
},s^{\prime })_{\phi _{j}}=(\omega ,\sigma ,p,s|_{\phi _{i}}|\omega ^{\prime
},\sigma ^{\prime })_{\phi _{i}}=0  \label{L.3.3}
\end{equation}
Then, a generic functional of $\widehat{{\cal A}^{\prime }}$ reads 
\[
\widehat{\rho }=\sum_{i}\int_{p\epsilon D_{\phi _{i}}}dp^{N}\int_{0}^{\infty
}\overline{\rho _{\phi _{i}}(\omega ,p,)}(\omega ,p|_{\phi _{i}}d\omega 
\]

\begin{equation}
+\sum_{i}\int_{p\epsilon D_{\phi _{i}}}\int_{p\epsilon D_{\phi
_{i}}}dp^{N}dp^{\prime N}\int_{0}^{\infty }\int_{0}^{\infty }\overline{\rho
_{\phi _{i}}(\omega ,\omega ^{\prime },p,p^{\prime })}(\omega ,\omega
^{\prime },p,p^{\prime }|_{\phi _{i}}d\omega d\omega ^{\prime }
\label{L.3.4}
\end{equation}
Like functions $O_{\phi _{i}}(\omega ,\omega ^{\prime },p,p^{\prime })$,
functions $\rho _{\phi _{i}}(\omega ,\omega ^{\prime },p,p^{\prime })$ are
regular and have all the mathematical properties necessary to make the
formalism successful (see \cite{Laura-Cast 1998-A}). Moreover, the $\widehat{%
\rho }$ must be self-adjoint, and their diagonal $\rho _{\phi _{i}}(\omega
,p)$ must represent probabilities; thus, $\sum_{i,p}\int_{0}^{\infty }\rho
_{\phi _{i}}(\omega ,p)d\omega =1$ (as in eq.(\ref{2.6'})) and, {\it most
important,} 
\begin{equation}
\rho _{\phi _{i}}(\omega ,p)\geq 0  \label{L.3.5}
\end{equation}
The $\widehat{\rho }$ with such properties belong to a convex set $\widehat{%
{\cal S}}$, the set of states. Also, as in eq.(\ref{1.9}), 
\[
(\widehat{\rho }|\widehat{O})=\sum_{i}\int_{p\epsilon D_{\phi
_{i}}}\int_{0}^{\infty }\overline{\rho _{\phi _{i}}(\omega ,p,)}O_{\phi
_{i}}(\omega ,p)d\omega dp^{N} 
\]

\begin{equation}
+\sum_{i}\int_{p\epsilon D_{\phi _{i}}}\int_{p\epsilon D_{\phi
_{i}}}\int_{0}^{\infty }\int_{0}^{\infty }\overline{\rho _{\phi _{i}}(\omega
,\omega ^{\prime },p,p^{\prime })}O_{\phi _{i}}(\omega ,\omega ^{\prime
},p,p^{\prime })d\omega d\omega ^{\prime }d^{N}p\,d^{N}p^{\prime }
\label{L.3.6}
\end{equation}
b.- Since $\widehat{{\cal L}}_{R}$ and $\widehat{{\cal L}_{R}^{\prime }}$
are spaces of operators on a Hilbert space ${\cal H}$, the symbol for any $%
\widehat{\rho }_{R}\in \widehat{{\cal L}_{R}^{\prime }}$ is defined as in
eq.(\ref{L.3.7}).\footnote{%
We repeat that, in the case of states, we must add a new factor $(2\pi \hbar
)^{-(N+1)}$ to definition (\ref{L.2.6}) in order to maintain the usual
normalization of $\rho (\phi )$.} From this definition, eq.(\ref{L.3.8}) can
be proved for the regular parts (see the demonstration in \cite{Wigner},
eq.(2.13)): 
\begin{equation}
(\widehat{\rho }_{R}|\widehat{O}_{R})=(symb\widehat{\rho }_{R}|symb\widehat{%
O_{R}})=\sum_{i}\int_{D_{\phi _{i}}}d\phi ^{2(N+1)}\rho _{\phi _{i}R}(\phi
)O_{\phi _{i}R}(\phi )  \label{L.3.8'}
\end{equation}
Then, in $\widehat{{\cal L}}_{R}$ and $\widehat{\text{ }{\cal L}_{R}^{\prime
}}$ all the equations are the usual ones (i.e. those of papers \cite{Wigner}
and \cite{Symb}).

Let us now consider the singular dual space $\widehat{{\cal L}_{S}^{\prime }}
$, the case not treated in the bibliography. In this space we will {\it %
define} $symb\widehat{\rho }_{S}$ as the function on ${\cal M}$ that
satisfies an equation similar to eqs.(\ref{L.3.8}) or (\ref{L.3.8'}) for any 
$\widehat{O}_{S}\in \widehat{{\cal L}}_{S}$, namely, 
\[
(\widehat{\rho }_{S}|\widehat{O}_{S})\doteq (symb\widehat{\rho }_{S}|symb%
\widehat{O_{S}}) 
\]
precisely,

\begin{equation}
\sum_{i}\int_{p\epsilon D_{\phi _{i}}}\int_{0}^{\infty }\rho _{\phi
i}(\omega ,p,)O_{\phi _{i}}(\omega ,p)d\omega dp^{N}=\sum_{i}\int_{D_{\phi
_{i}}}d\phi ^{2(N+1)}\rho _{\phi _{i}S}(\phi )O_{\phi _{i}S}(\phi )
\label{L.3.10}
\end{equation}
where the unknown density function $\rho _{S}(\phi )=symb\widehat{\rho }_{S}$
can be decomposed as 
\begin{equation}
symb\widehat{\rho }_{S}=\rho _{S}(\phi )=\sum_{i}\rho _{\phi _{i}S}(\phi )
\label{L.3.9'}
\end{equation}
in each $D_{\phi _{i}}$. Thus, since we know $\rho _{\phi i}(\omega ,p,)$, $%
O_{\phi _{i}}(\omega ,p)$, and $O_{\phi _{i}S}(\phi )$, we can compute $\rho
_{\phi _{i}S}(\phi )$ to obtain $\rho _{S}(\phi )=symb\widehat{\rho }_{S}$.
Now (as suggested by eq.(\ref{L.2.20})), $\widehat{\rho _{S}},$ being time
invariant, must be a function of the constants of the motion; therefore (as
in Subsection A) its Weyl-transformed $\rho _{S}(\phi )$ must be endowed
with the same property, but now in the classical case. Since the $\{H(\phi
),P_{\phi _{i}}(\phi )\}$ are locally a complete set of constants of the
motion in involution, we must have 
\begin{equation}
\rho _{\phi _{i}}(\phi )=F(H(\phi ),P_{\phi _{i}}(\phi ))  \label{L.3.11}
\end{equation}
We will find the function $F$. The system has a local pointer CSCO of $N+1$
operators and the dimension of its phase space is $2(N+1)$, i.e. it is {\it %
locally} an integrable system.\footnote{%
We have discussed this fact in detail at the beginning of section III. The
constants $J$ are global or isolating in the case of an integrable system,
but not in the non-integrable case. Nevertheless, they are locally defined.
Moreover, we will only consider the cases where action-angle variables can
be locally defined.} Then, we can define {\it locally at} $D_{\phi _{i}}$
the action angle variables ($\theta ^{0},\theta ^{1},...,\theta ^{N},J_{\phi
_{i}}^{0},J_{\phi _{i}}^{1},...,J_{\phi _{i}}^{N})$, where $J_{\phi
_{i}}^{0},$ $J_{\phi _{i}}^{1},...,J_{\phi _{i}}^{N}$ would be just $%
H,P_{\phi _{i}1},...,P_{\phi _{i}N}$ (multiplied by adequate constants in
such a way that the $\theta _{\phi _{i}}^{i0}$ variables belong to an
interval $0\leq \theta _{\phi _{i}}^{i0}\leq 2\pi $ in the integrable case).
We will call '$J$' just the '$H,P_{\phi _{i}}$'. Thus, we can make the
canonical transformation $\phi ^{a}\rightarrow \theta _{\phi
_{i}}^{0},\theta _{\phi _{i}}^{1},...,\theta _{\phi _{i}}^{N},H,P_{\phi
_{i}1},...,P_{\phi _{i}N}$, and we obtain 
\begin{equation}
d\phi ^{2(N+1)}=dq^{(N+1)}dp^{(N+1)}=d\theta _{\phi _{i}}^{(N+1)}dHdP_{\phi
_{i}}^{N}  \label{L.3.12}
\end{equation}
because the Jacobian of a canonical transformation is $\pm 1$. Since the
transformation must keep the metric of eq.(\ref{L.2.5}), we can take $1$\
with no loss of generality.

In order to compute the l.h.s. of eq.(\ref{L.3.10}), we must know how to
integrate functions $f(H,P_{\phi _{i}})=f(H,P_{\phi _{i}},...,P_{\phi _{i}})$
which are just functions of the constant of motion, precisely, 
\[
\int_{D_{\phi _{i}}}d\phi ^{2(N+1)}f(H,P_{\phi _{i}})=\int_{D_{\phi
_{i}}}d\theta _{\phi _{i}}^{(N+1)}dHdP_{\phi _{i}}^{N}f(H,P_{\phi _{i}}) 
\]
\begin{equation}
=\int_{D\phi _{i}}dHdP_{\phi _{i}}^{N}C_{\phi _{i}}(H,P_{\phi
_{i}})f(H,P_{\phi _{i}})  \label{L.3.13}
\end{equation}
where we have integrated the angular variables $\theta _{\phi
_{i}}^{0},\theta _{\phi _{i}}^{1},...,\theta _{\phi _{i}}^{N}$ and obtained
the configuration volume $C_{\phi _{i}}(H,P_{\phi _{i}})$ of the portion of
the hypersurface defined by $(H=const.,P_{\phi _{i}}=const.)$ and contained
in $D_{\phi _{i}}$. So, from eqs.(\ref{L.3.10}) and (\ref{L.3.13}) we have
that 
\[
\int_{p\epsilon D_{\phi _{i}}}\int_{0}^{\infty }\rho _{\phi i}(\omega
,p,)O_{\phi _{i}}(\omega ,p)d\omega dp^{N} 
\]
\begin{equation}
=\int dHdP_{\phi _{i}}^{N}C_{\phi _{i}}(H,P_{\phi _{i}})\rho _{\phi
_{i}S}(H,P_{\phi _{i}})O_{\phi _{i}S}(H,P_{\phi _{i}})  \label{l.15}
\end{equation}
for all $O_{\phi _{i}}(H,P_{\phi _{i}})=O_{S\phi _{i}}(H,P_{\phi _{i}})$
(see eq.(\ref{interp})). The last equation defines $\rho _{S\phi _{i}}(H,P)=%
\frac{1}{C_{\phi _{i}}}\rho _{\phi _{i}}(H,P)$ for $\phi \in {\cal D}_{\phi
_{i}}$,\footnote{%
In the integrable case, where there is just one $\rho (H,P)$, it would be $%
\rho _{\phi _{i}}(H,P)=\frac{C_{\phi _{i}}(H,P)}{2\pi ^{N+1}}\rho (H,P)$ and
the results of paper \cite{Cast-Laura 2000-PRA} would be reobtained. In
fact, integrating over a torus in the $\theta $ we have ($2\pi )^{N+1}\rho
(H,P)=\sum_{i}C_{\phi _{i}}(H,P)\rho _{\phi _{i}}(H,P)$.
\par
An example to fix the ideas: let us consider the harmonic oscillator and the
plane $q,p$ in radial coordinates $\theta ,H$. Let us define two $D\phi _{i}$%
: $D_{1}$ with $0\leq \theta <\Theta (H)$ and $D_{2}$ with $\Theta (H)\leq
\theta <2\pi $, where $\Theta (H)$ is an arbitrary function. Then, 
\[
\rho (\phi )=\rho _{1}(\phi )I_{1}(\phi )+\rho _{2}(\phi )I_{2}(\phi ) 
\]
If $\rho (\phi )=\rho (H)$, by integrating over the $\theta $ we obtain 
\[
2\pi \rho (H)=\int_{0}^{\Theta (H)}\rho _{1}(H)I_{1}(\phi )d\theta
+\int_{\Theta (H)}^{2\pi }\rho _{2}(H)I_{2}(\phi )d\theta 
\]
\[
=\rho _{1}(H)\Theta (H)+(2\pi -\Theta (H))\rho _{2}(H)= 
\]
\[
\rho _{1}(H)C_{1}(\phi )+\rho _{2}(H)C_{2}(\phi ) 
\]
namely, the equation $(2\pi )^{N+1}\rho (H,P)=\sum_{i}C_{\phi _{i}}(H,P)\rho
_{\phi _{i}}(H,P)$ for this particular case with dimension $N+1=1$} but not
for $\phi \in {\cal M}\backprime {\cal D}_{\phi _{i}}$\footnote{%
We will forget the joining zones ${\cal F}_{\phi _{i}}$\ and $F_{\phi _{i}}$%
\ since we have already proved that, when $S\rightarrow \infty $, their
influence is irrelevant.}; then, as in the case of $O_{S\phi _{i}}(\phi )$,
we will consider that $\rho _{S\phi _{i}}(\phi )=0$ for $\phi \in {\cal M}%
\backprime {\cal D}_{\phi _{i}}$ and that they are defined all over ${\cal M}
$ (this causes no problem because $O_{S\phi _{i}}(\phi )$ is multiplied by $%
\rho _{S\phi _{i}}(\phi )$, and $O_{S\phi _{i}}(\phi )$ has this property).
In this way, we can arrive from eq.(\ref{L.3.9'}) to our final result 
\begin{equation}
\rho _{S}(\phi )=\rho _{*}(\phi )=\sum_{i}\frac{1}{C_{\phi _{i}}(H,P_{\phi
_{i}})}\rho _{\phi _{i}}\left( H(\phi ),P_{\phi _{i}}(\phi )\right)
\label{1.17}
\end{equation}
Now, from eq.(\ref{L.3.5}) we obtain that 
\begin{equation}
\rho _{S}(\phi )=\rho _{*}(\phi )\geq 0  \label{1.17'}
\end{equation}
This means that the Wigner transformation of the singular part {\it can be
considered a density function since it is non-negatively defined} (of
course, this is not the case for the regular part).

Always working in the domain ${\cal D}_{\phi _{i}}$ and making $\rho _{\phi
_{i}}(\omega ,p)=\delta (\omega -\omega ^{\prime })\delta ^{N}(p-p^{\prime
}) $, we also have\footnote{%
In the chaotic, homogeneous, ergodic case, we have a $N+1-$CSCO with just $%
\widehat{H}$ and, classically, just $H$ as a constant of motion. In this
case (see \cite{Gutz}, p.247), 
\[
\rho _{S}(\phi )=\rho _{*}(\phi )=\int_{0}^{\infty }\delta (\omega -E)\frac{1%
}{C(H)}\delta (H(\phi )-E)=\frac{\delta (H(\phi )-E)}{\int dqdp\delta
(H(\phi )-E)} 
\]
} 
\begin{equation}
symb(\omega ^{\prime },p^{\prime },(\phi )|_{\phi _{i}}=\frac{1}{C_{\phi
_{i}}(H,P_{\phi _{i}})}\delta \left( H(\phi )-\omega ^{\prime }\right)
\delta ^{(N)}\left( P(\phi )-p_{\phi _{i}}^{\prime }\right)  \label{1.16}
\end{equation}

c.- From eqs.(\ref{WLP}) and (\ref{1.16}) we obtain

\begin{equation}
\rho _{S}(\phi )=\rho _{*}(\phi )=\sum_{i}\int_{p\epsilon D_{\phi
_{i}}}dp\int_{0}^{\infty }\rho _{\phi _{i}}(\omega ,p)\frac{1}{C_{\phi
_{i}}(H,P_{\phi _{i}})}\delta \left( H(\phi )-\omega \right) \delta
^{(N)}\left( P(\phi )-p_{\phi _{i}}\right) d\omega  \label{1.17''}
\end{equation}
The continuity of the function $\rho _{*}(\phi )$, when it goes from one $%
D_{\phi _{i}}$ to another $D_{\phi _{j}}$ ($i\neq j)$, is demonstrated in
Section V. Therefore, we have obtained a decomposition of $\rho _{*}(\phi )=$
$\rho _{S}(\phi )$ in classical hypersurfaces ($H=\omega ,$ $P_{\phi
_{i}}(\phi )=p_{\phi _{i}}$), containing classical trajectories, summed with
different positive weight coefficients $\rho _{\phi _{i}}(\omega ,p)/C_{\phi
_{i}}(H,P_{\phi _{i}})$, and represented in different ways in each domain $%
D_{\phi _{i}}$, but still with the same interpretation as in the integrable
case.\footnote{%
We can verify the normalization: 
\[
\int \rho _{S}(\phi )d\phi ^{2(N+1)}=\sum_{i}\int \rho _{S}(\phi )dHdP_{\phi
_{i}}^{N}d\theta _{\phi _{i}}^{N+1}= 
\]
\[
\sum_{i}\int dHdP_{\phi _{i}}^{N}\frac{\rho _{S}(H,P_{\phi _{i}})}{C_{\phi
_{i}}(H,P_{\phi _{i}})}\int d\theta _{\phi _{i}}^{N+1}= 
\]
\[
\sum_{i}\int dHdP_{\phi _{i}}^{N}\rho _{S}(H,P_{\phi _{i}})=1 
\]
}

d.- Since now we know how to deal with the singular part, we have defined
the mapping of the quantum space of states $\widehat{{\cal A}^{\prime }}$ on
the 'classical' space of states ${\cal A}^{\prime }$ 
\begin{equation}
symb:\widehat{{\cal A}^{\prime }}\rightarrow {\cal A}^{\prime }
\label{L.3.21}
\end{equation}
In the limit $\hbar \rightarrow 0$, eqs.(\ref{L.2.11}) and (\ref{L.2.12})
are always valid; then, it might be supposed that we have arrived to the
classical limit for the states. But {\it this is not so} because, in
general, even for $S$ very big (or $\hbar $ very small) the obtained $\rho
(\phi )$ {\it does not satisfy} the condition (see also the Appendix of ref. 
\cite{Cast-Gadella 2003}) 
\begin{equation}
\rho (\phi )\geq 0  \label{L.3.22}
\end{equation}
This is due to the fact that the regular part is still present and this part
does not satisfy the last condition (on the contrary, from eq.(\ref{1.17'})
we can see that the singular part satisfies the last inequality). As a
consequence, $\rho (\phi )$ is not a density function and, therefore, the
mapping (\ref{L.3.21}) is not a mapping of quantum mechanics on classical
statistical mechanics. This mapping does not give us the classical world,
but a deformed classical world where 'density functions' can be negative. In
other words, when $\hbar \rightarrow 0$ the isomorphism (\ref{L.3.21}) is a
mapping of quantum mechanics on a certain quantum mechanics 'alla classica',
namely, only formulated in phase space ${\cal M}$ but not satisfying eq. (%
\ref{L.3.22}). This clearly shows that $\hbar \rightarrow 0$ is not the
classical limit. In order to obtain this limit, we have to introduce
decoherence, as previously studied, both at the quantum and the classical
level.

\subsection{Time evolution and decoherence}

As we have seen, the only thing that prevents us from having a good
isomorphism (\ref{L.3.21}) is that the regular parts do not satisfy
condition (\ref{L.3.22}). But we know from eqs.(\ref{WLM}) or (\ref{WLP} )
that, for $t\rightarrow \infty $, the regular part vanishes and only the
singular part remains, which does satisfy this condition. As a consequence,
after decoherence and $\hbar \rightarrow 0$ (that is, the elimination of all
the $0(\hbar ^{2})$ that we have omitted), we finally obtain the classical
statistical limit since the classical densities obtained obey all the laws
of classical statistical mechanics. In fact, as we will see in the next
section in detail, eq.(\ref{1.17''}) shows that these distributions are the
result of classical point-like-states moving in phase space and following
classical trajectories. The usual classical limit is obtained by choosing
one of these trajectories; we will explain this procedure in the next
section.

\section{The classical limit}

From what we have learnt above, we can explain with more detail the three
steps involved in the classical limit, presented in the introduction and
shown in the following diagram: 
\[
Quantum\text{ }Mechanics-(decoherence)\rightarrow Boolean\text{ }Quantum%
\text{ }Mechanics-(symb\text{ }and\text{ }\hbar \rightarrow 0\text{ }%
)\rightarrow 
\]
\[
Classical\text{ }Statistical\text{ }Mechanics-(choice\text{ }of\text{ }a%
\text{ }trajectory)\rightarrow Classical\text{ }Mechanics 
\]
Let us comment these three steps:

i.- $Quantum$ $Mechanics-(decoherence)\rightarrow Boolean$ $Quantum$ $%
Mechanics$. Decoherence transforms non-Boolean quantum mechanics into
Boolean quantum mechanics\footnote{%
Namely, quantum mechanics in the local CSCO $\{\widehat{H},\widehat{P}\}$
using only diagonal states.} since it eliminates the off-diagonal terms, as
we have shown in eq.(\ref{WLP}).

ii.-$Boolean$ $Quantum$ $Mechanics-(symb$ $and$ $\hbar \rightarrow 0$ $%
)\rightarrow Classical$ $Statistical$ $Mechanics$. The Wigner transformation 
$symb$ and the limit $\hbar \rightarrow 0$ are defined with no problems in
the singular part remaining after decoherence. They map Boolean quantum
mechanics onto classical statistical mechanics: this is what we have
essentially shown above. Our demonstration culminates in Section IV.C, where
we have proved that the transformed quantum Boolean states are really
positively defined densities. From eq.(\ref{1.17''}) we also know that these
densities are the sums of densities strongly peaked on the classical
hypersurfaces defined by the constants of the motion $H(\phi )=\omega ,$ $%
P_{\phi _{i}}(\phi )=p_{\phi _{i}}$. In the next step we will see that such
classical hypersurfaces contain classical trajectories averaged by the
coefficients $\rho _{\phi _{i}}(\omega ,p)$.

iii.-$Classical$ $Statistical$ $Mechanics-(choice$ $of$ $a$ $%
trajectory)\rightarrow Classical$ $Mechanics$ ($Localization$ $or\
Actualization)$. After step (ii), we are still in classical statistical
mechanics but not in proper classical mechanics. To perform the last step we
have to pass from classical densities to classical trajectories (i.e. to
consider the localization effect\footnote{%
See \cite{Stockmann}, Chap. 4, for a different view.}). For this purpose let
us observe that, after the two first steps, the formalism of Boolean quantum
mechanics is isomorphic with the formalism of statistical classical
mechanics:

\begin{itemize}
\item  {\it For the observables}: After $symb$ and $\hbar \rightarrow 0$, we
obtain the correspondence $symb:\widehat{{\cal A}}\sim {\cal A}$ (see
Section IV A), namely, 
\[
A_{\phi _{i}}(\widehat{H,}\widehat{P}_{\phi _{i}})\sim A_{\phi _{i}}(H(\phi
),P_{\phi _{i}}(\phi )) 
\]

\item  {\it For the states}: After decoherence,{\it \ }$symb$ and $\hbar
\rightarrow 0$, again $symb:\widehat{{\cal A}^{\prime }}\sim {\cal A}%
^{\prime }$ (see Section IV.B), namely, 
\[
\rho _{\phi _{i}}(\widehat{H,}\widehat{P}_{\phi _{i}})\sim \rho _{\phi
_{i}}(H(\phi ),P_{\phi _{i}}(\phi ))\geq 0 
\]
\end{itemize}

and the states $\rho _{*}(\widehat{H,}\widehat{P})$ and $\rho _{*}(H(\phi
),P(\phi ))$ are time invariant: 
\[
\rho _{*}(\widehat{H,}\widehat{P})=\sum_{i}\int d\omega \int_{p\epsilon
D_{\phi _{i}}}dp^{N}\frac{\rho _{\phi _{i}}(\omega ,p)}{C(\omega ,p)}(\omega
,p|_{\phi _{i}} 
\]
\begin{equation}
\rho _{*}(H(\phi ),P(\phi ))=\sum_{i}\int d\omega \int_{p\epsilon D_{\phi
_{i}}}dp^{N}\frac{\rho _{\phi _{i}}(\omega ,p)}{C_{\phi _{i}}(\omega ,p)}%
\delta \left( H(\phi )-\omega \right) \delta ^{N}\left( P_{\phi _{i}}(\phi
)-p_{\phi _{i}}\right)  \label{5.1}
\end{equation}
Moreover, since $\Delta A\Delta B\geq \frac{\hbar }{2}|\langle [A,B]\rangle
_{\rho }|$, in the limit $\hbar \rightarrow 0$ there are no uncertainty
relations and the algebras $\widehat{{\cal A}}$ and ${\cal A}$ can be
considered commutative (remember that, according to the uncertainty
principle, $\hbar \rightarrow 0$ has the same effect that $[A,B]=0$). In
other words, in the limit $\hbar \rightarrow 0$ all the picture is classical
in such a way that the trajectories, contained in the hypersurfaces $H(\phi
)=\omega $, $P_{\phi _{i}}(\phi )=p_{\phi _{i}}$, could be interpreted as 
{\it real classical trajectories}. However, the $\delta (H(\phi )-\omega
)\delta ^{N}(P_{\phi _{i}}(\phi )-p_{\phi _{i}})$ still represent states
strongly peaked around these hypersurfaces (but not around trajectories).
Therefore, if we want to obtain an equation like (\ref{5.1}) but clearly
showing the classical trajectories, we have to introduce the initial
conditions of each trajectory.

Let us consider a classical trajectory in phase space ${\cal M=M}_{2(N+1)}$,
expressed in the classical coordinates $(\tau ,\theta _{\phi _{i}},H,P_{\phi
_{i}})$, where $\tau $ is the coordinate canonically conjugated to $H$ and
the $\theta _{\phi _{i}}$ are the coordinates canonically conjugated to the $%
P_{\phi _{i}}$. The constants of the motion in involution are $\{H,P_{\phi
_{i}}\}$; but, for conciseness and generality, let us consider that the
constants of the motion in involution are $\{\Pi _{\phi _{i}}\}$ with
conjugated coordinates $\{A_{\phi _{i}}\}$, and that $H=H(\Pi _{\phi _{i}})$%
. From the von Neumann-Liouville equation in the Heisenberg representation, 
\[
i\hbar \frac{d\widehat{A}}{dt}=[A,H] 
\]
we obtain 
\[
\frac{dA(\phi )}{dt}=\{A,H\}_{mb}=\{A,H\}_{pb}+0(\hbar ^{2}) 
\]
Then, the Hamiltonian equations in the limit $\hbar \rightarrow 0$ read%
\footnote{%
These equations correspond to the system of differential equations (3.1) of 
\cite{Mack}.} 
\begin{equation}
\frac{dA_{\phi _{i}}}{dt}=\frac{\partial H}{\partial \Pi _{\phi _{i}}}%
=\Omega _{\phi _{i}}(\Pi _{\phi _{i}})=const.;\qquad \frac{d\Pi _{\phi _{i}}%
}{dt}=-\frac{\partial H}{\partial A_{\phi _{i}}}=0  \label{V.2}
\end{equation}
The classical trajectories are 
\begin{equation}
A_{\phi _{i}}(t)=A_{\phi _{i}}^{(0)}+\Omega _{\phi _{i}}(\Pi _{\phi
_{i}})t,\qquad \Pi _{\phi _{i}}=\Pi _{\phi _{i}}^{(0)}=const.  \label{V.3}
\end{equation}
where the $A_{\phi _{i}}^{(0)}$ and $\Pi _{\phi _{i}}^{(0)}$ are integration
constants. A distribution strongly peaked on this trajectory reads 
\[
\delta [A_{\phi _{i}}(t)-A_{\phi _{i}}^{(0)}-\Omega _{\phi _{i}}(\Pi _{\phi
_{i}})t]\delta (\Pi _{\phi _{i}}-\Pi _{\phi _{i}}^{(0)}) 
\]
and a general classical distribution evolving according to the motion (\ref
{V.3}) reads\footnote{%
If the evolution $S^{t}$ of \cite{Mack} were the (\ref{V.3}), the
corresponding density would be $f(t,x)\equiv P^{t}f(x)$ (see \cite{Mack},
eq.(3.2)) where $P^{t}$ would represent a Frobenius-Perron evolution.
Moreover, it is easy to show that $\rho (t,\phi )$ satisfies the Liouville
equation.} 
\[
\rho _{C}(t,\phi )=\sum_{i}\int_{D_{\phi _{i}}}\rho _{\phi
_{i}}^{(C)}(A_{\phi _{i}}^{(0)},\Pi _{\phi _{i}}^{(0)})\delta [A_{\phi
_{i}}(t)-A_{\phi _{i}}^{(0)}-\Omega _{\phi _{i}}(\Pi _{\phi _{i}})t]\times 
\]
\begin{equation}
\delta (\Pi _{\phi _{i}}-\Pi _{\phi _{i}}^{(0)})d^{N+1}A_{\phi
_{i}}^{(0)}d^{N+1}\Pi _{\phi _{i}}^{(0)}  \label{88'}
\end{equation}
where $\rho _{\phi _{i}}^{(C)}(A_{\phi _{i}}^{(0)},\Pi _{\phi _{i}}^{(0)})$
is a generic classical coefficient (undefined up to now ). If we want that
this density (evolving according to a Frobenius-Perron evolution, see \cite
{Mack}) be an equilibrium density, we have to eliminate the variable $t$.
For this purpose, it is sufficient to choose the initial distribution $\rho
_{\phi _{i}}^{(C)}(A_{\phi _{i}}^{(0)},\Pi _{\phi _{i}}^{(0)})$ as just a
function of $\Pi _{\phi _{i}}^{(0)}$, namely, $\rho _{\phi _{i}}^{(C)}(\Pi
_{\phi _{i}}^{(0)})$, which is still a free function that we can use to
represent different $\rho _{C}(\phi )$. Then, we obtain 
\begin{equation}
\rho _{C}(\phi )=\sum_{i}\int_{D_{\phi _{i}}}\rho _{\phi _{i}}^{(C)}(\Pi
_{\phi _{i}}^{(0)})\delta (\Pi _{\phi _{i}}-\Pi _{\phi
_{i}}^{(0)})d^{N+1}\Pi _{\phi _{i}}^{(0)}  \label{V.4}
\end{equation}
since, for any fixed $t$, we have 
\[
\sum_{i}\int_{D_{\phi _{i}}}\delta (A_{\phi _{i}}(t)-A_{\phi
_{i}}^{(0)}-\Omega _{\phi _{i}}(\Pi _{\phi _{i}})t)d^{N+1}A_{\phi
_{i}}^{(0)}=1 
\]
Going back to our primitive variables, eq.(\ref{88'}) reads 
\[
\rho _{C}(\phi )=\sum_{i}\int \rho _{\phi _{i}}^{(C)}(\omega ,p)\delta
(H(\phi )-\omega )\delta (P_{\phi _{i}}-p_{\phi _{i}}) 
\]
\begin{equation}
\times \delta (\tau (\phi )-\tau _{0}-\omega t)\delta (\theta _{\phi
_{i}}(\phi )-\theta _{\phi _{i}0}-p_{\phi _{i}}t)d\omega d^{N}p_{\phi
_{i}}d\tau _{0}d\theta _{\phi _{i}0}  \label{V.5}
\end{equation}
while eq.(\ref{V.4}) reads 
\[
\rho _{C}(\phi )=\sum_{i}\int \rho _{\phi _{i}}^{(C)}(\omega ,p)\delta
(H(\phi )-\omega )\delta (P_{\phi _{i}}-p_{\phi _{i}})d\omega d^{N}p_{\phi
_{i}} 
\]
Then, from eq.(\ref{5.1}) and making the undefined coefficient $\rho _{\phi
_{i}}^{(C)}(\omega ,p)=\frac{\rho _{\phi _{i}}(\omega ,p)}{C_{\phi
_{i}}(\omega ,p_{\phi _{i}})}$, we have 
\begin{equation}
\rho _{*}(\phi )=\rho _{C}(\phi )  \label{V.6}
\end{equation}
The function $\rho _{C}(\phi )$ can be interpreted as the equilibrium
density of a Frobenius-Perron evolution of particle-like states $(\tau
,\theta _{\phi _{i}},H,P_{\phi _{i}})$, {\it as if} these states would move
in phase space ${\cal M=M}_{2(N+1)}$ according to the classical motions (\ref
{V.3}).

However, each term of the sum $\sum_{i}$ of eq.(\ref{V.5}) is valid in the
chart ${\cal D}_{\phi _{i}}$ ($D_{\phi _{i}}\subset {\cal D}_{\phi _{i}}$).
In a different chart ${\cal D}_{\phi j}$ ($D_{\phi _{j}}$ $\subset {\cal D}%
_{\phi j}$), the equation is also valid and, then, at $\phi \in {\cal D}%
_{\phi _{i}}\cap {\cal D}_{\phi _{j}}$ we have 
\[
\int \frac{\rho _{\phi _{i}}(\omega ,p)}{C_{\phi _{i}}(\omega ,p_{\phi _{i}})%
}\delta (H(\phi )-\omega )\delta (P_{\phi _{i}}-p_{\phi _{i}}) 
\]
\[
\times \delta (\tau (\phi )-\tau _{0}-\omega t)\delta (\theta _{\phi
_{i}}(\phi )-\theta _{\phi _{i}0}-p_{\phi _{i}}t)d\omega d^{N}p_{\phi
_{i}}d\tau _{0}dA_{\phi _{i}0}= 
\]
\[
\int \frac{\rho _{\phi _{i}}(\omega ,p)}{C_{\phi _{i}}(\omega ,p_{\phi _{j}})%
}\delta (H(\phi )-\omega )\delta (P_{\phi _{j}}-p_{\phi _{j}}) 
\]
\[
\times \delta (\tau (\phi )-\tau _{0}-\omega t)\delta (\theta _{\phi
_{j}}(\phi )-\theta _{\phi _{j}0}-p_{\phi _{j}}t)d\omega d^{N}p_{\phi
_{j}}d\tau _{0}dA_{\phi _{j}0} 
\]
Here it is worth emphasizing that the trajectories $H=\omega $, $P_{\phi
_{i}}(\phi )=p_{\phi _{i}}$, $\tau (\phi )=\tau _{0}+\omega t,$ $\theta
_{\phi _{i}}(\phi )=\theta _{\phi _{i}0}+p_{\phi _{i}}t$ in chart ${\cal D}%
_{\phi _{i}}$ are {\it continuously connected} with those $H=\omega $, $%
P_{\phi _{j}}(\phi )=p_{\phi _{j}}$, $\tau (\phi )=\tau _{0}+\omega t,$ $%
\theta _{\phi _{j}}(\phi )=\theta _{\phi _{j}0}+p_{\phi _{j}}t$ in chart $%
{\cal D}_{\phi _{j}}$, because these charts are not generic but constructed
using the solution of eqs.(\ref{1.3}), (\ref{M.2}) or (\ref{V.2}). Since $%
D_{\phi _{i}}\subset {\cal D}_{\phi _{i}}$ and $D_{\phi _{j}\text{ }}\subset 
{\cal D}_{\phi _{j}}$, the same holds for the trajectories going from $%
D_{\phi _{i}}$ to $D_{\phi _{j}}$. Thus, the continuous connection follows
from the fact that {\it one and only one} solution of the trajectory
equation passes for each point of ${\cal M}$ (and, therefore, for each $\phi
\in {\cal D}_{\phi _{i}}\cap {\cal D}_{\phi _{j}}$).

Summing up, we have obtained a decomposition of $\rho _{*}(\phi )=$ $\rho
_{S}(\phi )$ in classical trajectories $H=\omega $, $P_{\phi _{i}}(\phi
)=p_{\phi _{i}}$, $\tau (\phi )=\tau _{0}+\omega t$, $\theta _{\phi
_{j}}(\phi )=\theta _{\phi _{j}0}+p_{\phi _{j}}t$, summed with different
weight coefficients $\rho _{\phi _{i}}(\omega ,p)/C_{\phi _{i}}(H,P_{\phi
_{i}})$ and represented in different ways in each domain ${\cal D}_{\phi
_{i}}$, but still with the same interpretation as in the integrable case.
Moreover, as announced in Section III.A.c, we see that chart ${\cal D}_{\phi
_{i}}$ is continuously connected with chart ${\cal D}_{\phi _{j}}$, for any $%
{\cal D}_{\phi _{i}}$, ${\cal D}_{\phi _{j}}$. Therefore, we have finally
obtained the classical limit to the extent that we have described each one
of the classical trajectories. But, since from the very beginning our system
was a non-integrable one, we have obtained {\it the classical limit of a
non-integrable system}, where the tori are broken and the trajectories are 
{\it potentially chaotic trajectories}.

Finally, we must remark that:

\begin{itemize}
\item  Each one of the described processes, decoherence, route to
macroscopicity, i.e. $\hbar \rightarrow 0$ (e.g. the macroscopicity obtained
when the two rays of an Stern-Gerlach experiment gradually separate), and
eventually localization (e.g. by a localizing potential, see \cite
{Cast-Laura 2000-PRA}, Appendix A), has its own characteristic time; in
particular, the decoherence time is computed in \cite{Cast-Lombardi 2005}.

\item  We have explained the classical limit as if each process
(decoherence, macroscopicity, and localization) took place one after the
other, only for didactical reasons. But this is an oversimplified picture of
the phenomenon. In fact, this may be not the case if the different processes
overlap. Considering that they have different characteristic times, there
are different possibilities according to the order in which the processes
finish.
\end{itemize}

\section{Partially non-integrable systems: the trace of the non-integrable
part}

In this section we will consider some well-known definitions, find some
relations with the concepts of statistical physics, and discuss the trace of
the non-integrable part of the system. All the definitions are formulated in
the classical language, but they become quantum definitions if we translate
them with the Weyl-Wigner-Moyal isomorphism. In some sense, that we will
precise elsewhere, these quantum definitions correspond to a quantum chaotic
hierarchy.

The classical constants of the motion $H(\phi ),$ $P_{1}(\phi ),...,$ $%
P_{N}(\phi )$ can be rigorously classified as \cite{Arnold}, \cite{Tabor}, 
\cite{Balescu}:

i.- {\bf Global or isolating constants of the motion, }that we will call '$H$%
', 
\begin{equation}
H_{0}(\phi )=H(\phi ),\text{ }H_{1}(\phi )=P_{1}(\phi ),...,H_{A}(\phi
)=P_{A}(\phi )  \label{3.1}
\end{equation}
The constants of the motion are global when the conditions 
\begin{equation}
H_{\alpha }=p_{\alpha },\qquad \alpha =0,...,A  \label{3.3}
\end{equation}
define, for each set of constants $(p_{0},...,p_{A})$, a global sub-manifold 
${\cal M}(p_{0},...,p_{A})$ (a torus in the bounded case) of phase space
where the trajectories are confined.\footnote{%
Isolating constants of the motion correspond to the 'simple' constants of
the motion in \cite{Katz}, p.60.} The dimension of ${\cal M}$ $%
(p_{0},...,p_{A})$ is $2(N+1)-(A+1)=2N$ $-A+1$.

{\bf ii.- Local or non-isolating constants of the motion, }that we will call
'$J_{\phi _{i}}$', 
\begin{equation}
J_{\phi _{i}1}(\phi )=P_{\phi _{i}A+1}(\phi ),...,J_{\phi _{i}N-A}(\phi
)=P_{\phi _{i}N}(\phi )  \nonumber
\end{equation}
The constants of the motion are local when the conditions 
\begin{equation}
J_{\phi _{i}\beta }=p_{\phi _{i}\beta +A}\qquad \beta =1,..,N-A  \label{3.4}
\end{equation}
do {\it not} define any global sub-manifold, but define just a local
sub-manifold at $D_{\phi _{i}}$.\footnote{%
Let us list the introduced dimensions:
\par
i.- The total dimension is $2(N+1).$%
\par
ii.- The number of the isolating constants is $A+1.$%
\par
iii.- The number of the non-isolating constants is $N-A.$%
\par
iv.- The number of the configuration coordinates is $N+1.$%
\par
v.- The dimension of ${\cal M}(p_{0},...,p_{A})$ is $2N-A+1$} The $J_{\phi
_{i}\beta }$ are {\it local momentum coordinates}.

When $A=N$, we say that the system is {\it integrable;} when $A=0$ we say
that the system is {\it non-integrable }(even if we will always consider $H$
globally defined), if $A<N$ we will say that the system is {\it partially
non-integrable}. Let us consider these three cases in detail.

\subsection{Integrable systems}

In this case, all the $P$ are isolating constants of the motion $H$, and
there are no $J_{\phi _{i}}$. Then, condition (\ref{3.3}) foliates the phase
space with submanifolds ${\cal M}(p_{0},...,p_{N})$ of dimension $N+1$,
labelled by the constants $(p_{0},...,p_{N})$. These submanifolds are tori
when the system is endowed with action-angle variables (e.g. when phase
space is bounded). On these tori, the motion of the configuration variables
is the motion described by classical mechanics. In the generic case, the
frequencies of the motions are {\it not rationally dependent} (or
non-commensurable); thus, the corresponding trajectories fill each torus in
a dense way: the motion is {\it ergodic} on each torus.\footnote{%
If the system is unbounded, ergodicity requires that the trajectories be
dense in a domain ${\cal D}(p_{0},...,p_{N})\subset {\cal M}%
(p_{0},...,p_{N}) $ (see subsection B).} Moreover, we can see from eq.(\ref
{1.17}) or in paper \cite{Cast-Laura 2000-PRA} that, if the angle-action
variables exist, there is a unique equilibrium state on each torus, 
\begin{equation}
\rho _{*}(\phi )=\frac{1}{(2\pi )^{N+1}}\rho (p_{0},p_{1},...,p_{N})
\label{3.4'}
\end{equation}
which is constant in the submanifold ${\cal M}(p_{0},...,p_{N})$; this
corresponds to a{\it \ microcanonical} equilibrium on each torus.

\subsection{Partially non-integrable systems}

In this case, not all the $P$ are isolating constants of the motion; so,
there are $H$ and $J_{\phi _{i}}.$ Then, the trajectories must be dense in a
domain ${\cal D}(p_{0},...,p_{A})\subset {\cal M}(p_{0},...,p_{A})$ of
dimension $2N-A+1$. In fact, if the dimension of ${\cal D}(p_{0},...,p_{A})$
were $<2N-A+1$, at least a new global constant would exist and there would
be $A+2$ global constants; but this is impossible since,{\bf \ }by its own
definition, $A+1$ is the total number of these constants. It is quite clear
that something as a 'thermodynamic equilibrium' classical density must be
globally defined in ${\cal D}(p_{0},...,p_{A})$ and, therefore, the $%
J_{1\phi _{0}}(\phi ),...,J_{N-A\phi _{0}}(\phi )$ cannot be explicit
variables{\it \ }of $\rho _{*}(\phi )$;\footnote{%
Even if another kind of equilibrium could be defined using local
coordinates, it is quite clear that only globally defined variables can play
the role of thermodynamic variables in eq.(\ref{3.9}).} then, we must have
(as in eq.(\ref{1.17})\footnote{%
There are not indices $\phi _{i}$ because the variables are the constants of
the motion globally defined in a manifold of $2(A+1)$ dimensions.}) 
\begin{equation}
\rho _{*}^{T}(\phi )=\frac{1}{C(H_{0}(\phi ),\text{ }H_{1}(\phi
),...,H_{A}(\phi ))}\rho (H_{0}(\phi ),\text{ }H_{1}(\phi ),...,H_{A}(\phi ))
\label{3.7}
\end{equation}
Since ${\cal D}(p_{0},...,p_{A})$ is contained in the hypersurface defined
by 
\begin{equation}
H_{0}(\phi )=p_{0}=const.,\text{\quad }H_{1}(\phi )=p_{1}=const.,\quad
...,\quad H_{A}(\phi )=p_{A}=const.  \label{3.8}
\end{equation}
$C(H_{0}(\phi ),$ $H_{1}(\phi ),...,H_{A}(\phi ))=C(p_{0},$ $%
p_{1},...,p_{A}) $ is the volume of the portion of the just mentioned
hypersurface contained in ${\cal D}(p_{0},...,p_{A})$ (in such a way that $%
\rho _{*}^{T}(\phi )$ would be normalized). Therefore,

\begin{equation}
\rho _{*}^{T}(\phi )=\frac{1}{C(p_{0},p_{1},...,p_{A})}\rho (p_{0},\text{ }%
p_{1},...,p_{A})=const.  \label{3.9}
\end{equation}
and we have found a unique equilibrium for each ${\cal D}(p_{0},...,p_{A})$,
namely, for each set of isolating constants of the motion or state variables 
$(p_{0},...,p_{A})$. Now we can repeat the argument of the integrable case.
In this case, phase space is foliated in submanifolds ${\cal M}%
(p_{0},...,p_{A})$ of dimension $2N-A+1$. The difference is that now not all
the coordinates of these submanifolds are configuration variables:\footnote{%
Regarding the configuration variables, it is clear that, in the
non-integrable case, none of them is a global constant of the motion. This
fact further reduces the dimension of ${\cal M}$ (or ${\cal D)}$. In fact:
\par
i.- The preserved tori satisfy an irrationality condition (\cite{Tabor}, eq.
(3.4.12)); so, the corresponding ratios of the frequencies are irrational
and the trajectories are dense in those tori.
\par
ii.- In the broken tori the trajectories are chaotic. Nevertheless, if in a
particular case there is a configuration variable $X$ that turns out to be a
global constant of the motion, it can be considered among the '$H$'. Then,
we will essentially work in the manifold $X=const.$ and nothing will change.}
since there are also momentum variables, in this case we cannot use the
argument about frequencies not rationally related. Nevertheless, in each $%
{\cal D}(p_{0},...,p_{A})\subset {\cal M}(p_{0},...,p_{A})$ there is a
unique equilibrium state (\ref{3.9}); as a consequence, by using theorem
(4.3) of ref. \cite{Mack} we can conclude that the motion is {\it ergodic.}%
\footnote{%
Theorem (4.3) requires that the evolution be a Frobenius-Perron one. As
explained in section V, this is precisely the case. In the equilibrium $%
\hbar \rightarrow 0$ limit, the motion takes place along classical
trajectories. So, even if $\rho _{*}(\phi )$ is constant in time, the
particles of the system are in motion.}

At this point, we may ask ourselves why the density $\rho _{*}(H_{0}(\phi
),H_{1}(\phi ),...,H_{A}(\phi )J_{1}(\phi ),...,J_{N-A}(\phi ))$ looses its $%
J_{\phi _{i}}$ variables. The physical reason for this relies on the fact
that the space of observables $\widehat{{\cal O}}$ contains only physical
measurable observables. In the integrable case, we have a global $N+1-$CSCO
whose observables can be globally measured in an independent way because
they commute. More generally, we can measure only the variables that belong
to a global set of commuting observables, even if the number of observables
is $<N+1$. In fact, the $J$ could be measured in a local $N+1-$CSCO, but
they change when they go from one local $N+1-$CSCO to another; therefore,
the period for making the measurement could be very short (eventually
shorter that the period necessary for the measurement itself, turning the
measurement impossible). In other words, since the classical momenta $%
J_{1\phi _{i}}(\phi ),...,J_{N-A\phi i}(\phi )$ have an ergodic motion, it
is reasonable to suppose that the quantum analogues $\widehat{J_{1\phi i}}%
,...,\widehat{J_{N-A\phi _{i}}}$ cannot be really measured, even at the
quantum level.\footnote{%
We can only measure dynamical variables when they can be considered as
constants in time, at least in the period of measurement. If a constant is
not global, it is only constant in time in a local coordinate system, i.e.
it is not a usual 'physical constant'.} This means that the set $\{\widehat{%
H_{0}},...,\widehat{H_{A}}\}$ is the relevant measurable global $A+1-$CSCO:
it is possible to measure only the $\widehat{H},\widehat{H_{1}},...,\widehat{%
H_{A}}$. Then, the isolating constants $H_{0}(\phi ),...,H_{A}(\phi )$ turn
out to be the only reliable characters in the quantum or the classical play,
and the only candidates for thermodynamic variables.

This point can be made in a different way. In the coordinates $J_{\phi _{i}}$%
, states can only be {\it locally diagonalized.} For this reason, it is
convenient to consider that the unitary operator $U$ of eq.(\ref{2.11'})
diagonalizes only the indices of the $H$ (that we will call $r$) and does
not diagonalize the indices of the $J$ (that we will call $m$). Then, we
obtain a basis where the coordinates of the states read (see (\ref{nonint})) 
\begin{equation}
\rho (\omega )_{\phi _{i}rmr^{\prime }m^{\prime }}=\rho _{\phi
_{i}rmm^{\prime }}(\omega )\delta _{rr^{\prime }}  \label{3.10}
\end{equation}
With this basis, eqs.(\ref{1.12}) and (\ref{1.13}) become 
\begin{equation}
\widehat{\rho _{*}}=W\lim_{t\rightarrow \infty }\widehat{\rho (t}%
)=\sum_{irmm^{\prime }}\int d\omega \rho _{\phi _{i}rmm^{\prime }}(\omega
)(\omega ,rmrm^{\prime }|_{\phi _{i}}  \label{3.11}
\end{equation}
and 
\begin{equation}
\widehat{P_{I}}=\sum_{irmm^{\prime }}\int d\omega P_{\phi _{i}rmm^{\prime
}}^{I}|\omega rm\rangle _{\phi _{i}}\langle \omega rm^{\prime }|_{\phi
_{i}},\qquad I=1,2,...,A  \label{3.12}
\end{equation}
and so on for the rest of the equations. But the $m$ indices can be 'traced
away' because the corresponding observables cannot be measured, i.e. they
cannot be considered classical in a global way. This strategy amounts to
consider that the $J$ operators and the $m$ indices are inexistent in space $%
\widehat{{\cal O}}$, to the extent that $\widehat{{\cal O}}$ is the space of
all measurable observables. This fact is manifested by the following change
in the observables of eq.(\ref{1.7'}) 
\begin{equation}
O(\omega )_{\phi _{0}rmr^{\prime }m^{\prime }}\rightarrow O(\omega )_{\phi
_{0}rr^{\prime }}\delta _{mm^{\prime }}  \label{3.13}
\end{equation}
where we only consider the diagonal term since we are only concerned with
the classical part. As we can see, the $m$ indices have a 'spherical
symmetry' and they measure nothing. Moreover, with variables $H$ and their
conjugate configuration variables, we can define a $2(A+1)$ manifold $\phi
_{0}$, and with variables $J$ and their conjugated configuration variables,
a $2(N-A)$ manifold $\phi _{irre}$, in such a way that ${\cal M=\phi }%
_{0}\otimes \phi _{irre}.$ Then, $O(\omega )_{\phi _{0}rr^{\prime }}$ is
globally defined in $\phi _{0}$ and 
\[
\widehat{O}_{S}=\sum_{irmr^{\prime }m^{\prime }}\int d\omega O(\omega
)_{\phi _{0}rr^{\prime }}\delta _{mm^{\prime }}|\omega rmr^{\prime
}m^{\prime })_{\phi _{i}}= 
\]
\[
\sum_{rr^{\prime }}\int d\omega O(\omega )_{\phi _{0}rr^{\prime
}}\sum_{im}|\omega rmr^{\prime }m)_{\phi _{i}}=\sum_{rr^{\prime }}\int
d\omega O(\omega )_{\phi _{0}rr^{\prime }}|\omega rr^{\prime })_{\phi _{0}} 
\]
where 
\[
|\omega rr^{\prime })_{\phi _{0}}=\sum_{im}|\omega rmr^{\prime }m)_{\phi
_{i}} 
\]
is the basis of the new space of observables. Then, the relevant part of eq.(%
\ref{1.7'}) reads 
\[
\langle \widehat{O}\rangle \widehat{_{\rho _{*}^{T}}}=\sum_{irmr^{\prime
}m^{\prime }}\int d\omega \overline{\rho (\omega )_{\phi _{i}rmr^{\prime
}m^{\prime }}}O(\omega )_{\phi _{i}rmr^{\prime }m^{\prime
}}=\sum_{irmr^{\prime }m^{\prime }}\int d\omega \overline{\rho _{\phi
_{i}rmm^{\prime }}(\omega )\delta _{rr^{\prime }}}O(\omega )_{\phi
_{0}rr^{\prime }}\delta _{mm^{\prime }}= 
\]
\begin{equation}
\sum_{r}\int d\omega \left( \sum_{im}\overline{\rho (\omega )_{\phi _{i}rmm}}%
\right) O(\omega )_{\phi _{0}rr}  \label{3.14}
\end{equation}
Calling $\sum_{im}\overline{\rho (\omega )_{\phi _{i}rmm}}=\overline{\rho
(\omega )_{\phi _{0}r}\text{ }}$, namely, making the $m-$trace of $\rho
(\omega )_{\phi _{0}rmr^{\prime }m}$, we obtain the 'traced' equation 
\begin{equation}
\langle \widehat{O}\rangle _{\rho _{*}^{T}}=\sum_{r}\int d\omega \overline{%
\rho (\omega )_{\phi _{0}r}\text{ }}O(\omega )_{\phi _{0}rr}  \label{3.15}
\end{equation}
where the $m$ indices have disappeared; eq.(\ref{3.11}) now reads

\[
\widehat{\rho _{*}}^{T}=W\lim_{t\rightarrow \infty }\widehat{\rho (t}%
)^{T}=\sum_{r}\int d\omega \rho (\omega )_{\phi _{0}r}(\omega ,r|_{\phi
_{0}} 
\]
where $T$ now means '$m-$traced' and $\{(\omega ,r|_{\phi _{0}}\}$ is the
dual basis of $\{|\omega ,r)_{\phi _{0}}\}$. If we work only with the $r$
indices $(r_{0},...,r_{A})$, the $\rho $ of eq.(\ref{3.4'}) becomes the $%
\rho $ of eq.(\ref{3.9}) solving the problem.\footnote{%
At this point, we could say that we are working with a {\it system} given by
the $\widehat{H_{0}},\widehat{H_{1}},...,\widehat{H_{A}}$ and an {\it %
environment} given by the $\widehat{J_{\phi _{0}1}},...,\widehat{J_{N-A\phi
_{0}}}$: we could consider this situation as the case of an {\it open
quantum system.}} After this coarse-graining that retains only the $r$
variables, the relevant phase space $\phi _{0}$ has only $2(A+1)$
dimensions. So, we can repeat all the theory with only these dimensions and
find that, at each manifold ${\cal M}(p_{0},p_{1},...,p_{A})$, there is only
one constant equilibrium state $\rho _{*}^{T}(\phi )$ given by eq.(\ref{3.9}%
) with $\rho (p_{0},p_{1},...,p_{A})=\rho (\omega )_{\phi _{0}r}$ ($\omega
=p_{0}$, $r=(p_{1},...,p_{A})$).

The above conclusions are important for two reasons:

1.- They prove that a system with a continuous evolution spectrum is ergodic
in ${\cal M}(p_{0},p_{1},...,p_{A})$ with a constant equilibrium density $%
\rho _{*}^{T}$. Therefore, such a system satisfies the two basic hypotheses
of classical statistical mechanics: it is {\it ergodic} and {\it %
microcanonical}.

2.- All the above classical properties can be transferred to the quantum
level; therefore, we can define {\it quantum ergodic systems} as those that
satisfy these transferred properties. This definition would be one of the
possible definitions of quantum chaos (see \cite{Earman}): precisely the one
which says that a quantum system is chaotic if it has a chaotic classical
limit (\cite{Naka}, \cite{Stockmann}). This fact is a consequence of our
definition of the classical limit.

\subsection{Non-integrable systems}

Let us now consider the particular case of a non-integrable system where
only $H(\phi )$ is globally defined ($A=0)$. Then, we can trace away all the 
$J_{\phi _{i}}$: the resulting traced system has a global CSCO $\{\widehat{H}%
\}$. As a consequence, in each energy manifold $H=\omega =const.$ the motion
is ergodic: this is the typical ergodic motion (the drop in the glass of
water, etc.). In this case, after tracing away the $J_{\phi _{i}}$ (i.e. the 
$r$-coordinates disappear), the pointer basis only depends on $H$, namely,
it is just the eigenbasis of $H$.

\section{Conclusions}

We want to conclude the paper proposing some suggestions for future research.

i.- We have essentially presented a {\it minimal formalism for quantum chaos}%
, to the extent that our quantum formalism satisfies a minimal requirement
for such a theory: by definition, a quantum chaotic system has, at least, a
classical non-integrable system as its classical limit. In fact, this is a
necessary but not a sufficient condition that any proposed theory of quantum
chaos must fulfil. Therefore, our next task is to address the question of
whether the set of phenomena known under the name of 'quantum chaos'(\cite
{Naka}, \cite{Stockmann}, \cite{Benatti}, \cite{Gutz},\cite{Earman}) can be
explained by means of our theoretical structure.

ii.- Quantum contexts are clearly related with $N+1-$CSCOs. We have seen
that generic $N+1-$CSCOs are local. This might have a relation with well
known physical questions, as the EPR problem and the Kochen-Speker and Bell
theorems (see \cite{Ballentine}), where paradoxes arise when we try to
describe the quantum system with just one CSCO.

iii.- In some sense, the equations of quantum physics have a local character
(\cite{Haag}, \cite{BLT}, \cite{Landau}); we have found that this is also
the case of the CSCOs: it might be useful to explore this analogy.

iv.- In quantum gravity we have to combine concepts of a local classical
theory, general relativity, with those of a global one, quantum mechanics,
and this task has shown to be almost impossible. Now that we have understood
the local nature of quantum mechanics, the task might become easier. Or
rephrased in a more modest way: in quantum field theory in curved space
time, we have the unsolved vacuum definition problem (\cite{Calzetta-Cast}, 
\cite{Cast-Mazzitelli}, \cite{Cast-Ferraro}), where the vacuum (a clearly
local classical general relativity notion) is considered on the global
grounds of the quantum field theory. Perhaps the present technique might
contribute to solve the problem.

\section{Acknowledgments.}

The authors are very grateful to Roland Omn\'{e}s for several interesting
comments, criticisms, and general informations. This paper was partially
supported by grants of the Buenos Aires University, the National Research
Council (CONICET), and the National Research Agency (FONCYT) of Argentina.

\appendix

\section{Integrability of the differential equations (\ref{1.3}) and (\ref
{M.2})}

a.- Let us consider eq.(\ref{1.3}), first as a mathematical partial
differential equation, and then from a physical point of view. Calling $%
\frac{\partial H}{\partial p_{qj}}=A_{j},$ $\frac{\partial H}{\partial q_{j}}%
=-B_{j}$, eq.(\ref{1.3}) reads 
\begin{equation}
\sum_{j=1}^{N}A_{j}\frac{\partial O_{I}}{\partial q_{j}}+B_{j}\frac{\partial
O_{I}}{\partial p_{qj}}=0  \label{Ap.1}
\end{equation}
where functions $A_{j},$ $B_{j}$ are known. To solve this equation around a
hypersurface ${\cal D}$ (containing a point $\phi _{i}$, that we will call $%
(q_{0}^{(0)},\varphi ^{(0)})$) is equivalent to find the trajectories
crossing ${\cal D}$ such that 
\begin{equation}
\frac{dq_{0}}{A_{0}}=...=\frac{dq_{N}}{A_{N}}=\frac{dp_{q0}}{B_{0}}=...=%
\frac{dp_{qN}}{B_{N}}  \label{Ap.2}
\end{equation}
This requires to solve the system of ordinary differential equations 
\[
\frac{dq_{1}}{dq_{0}}=\frac{A_{1}}{A_{0}}=a_{1},...,\frac{dq_{N}}{dq_{0}}=%
\frac{A_{N}}{A_{0}}=a_{N} 
\]
\begin{equation}
\frac{dp_{q0}}{dq_{0}}=\frac{B_{0}}{A_{0}}=b_{0},...,\frac{dp_{qN}}{dq_{0}}=%
\frac{B_{N}}{A_{0}}=b_{N}  \label{Ap.3}
\end{equation}
where we have taken $q_{0}$ as a parameter. Calling $\varphi =\{\varphi
_{\alpha }\}=\{q_{1},...,q_{N},p_{q0},...,p_{qN}\}$, the Lipschitz
conditions on functions $a$ and $b$ are satisfied if there exists a number $%
M>0$ such that 
\[
|a(q_{0},\varphi ^{\prime })-a(q_{0},\varphi )|\leq M\sum_{\alpha
=1}^{2N+1}|\varphi _{\alpha }^{\prime }-\varphi _{\alpha }| 
\]
\begin{equation}
|b(q_{0},\varphi ^{\prime })-b(q_{0},\varphi )|\leq M\sum_{\alpha
=1}^{2N+1}|\varphi _{\alpha }^{\prime }-\varphi _{\alpha }|  \label{Ap.4}
\end{equation}
If functions $a$ and $b$ are continuous and satisfy these conditions in a
neighborhood of a point $(q_{0}^{(0)},\varphi ^{(0)})$ considered as the
initial conditions, then there exists a unique solution of the system (\ref
{Ap.3}) in a neighborhood of $\varphi ^{(0)}$ and for $q_{0}\in
[q_{0}^{(0)}-c;q_{0}^{(0)}+c]$ for some $c>0$. Moreover, if also 
\[
\Delta =\left| 
\begin{array}{ll}
A_{j} & B_{j} \\ 
\frac{\partial q_{j}}{\partial \Phi _{j}} & \frac{\partial p_{j}}{\partial
\Phi _{j}}
\end{array}
\right| \neq 0 
\]
where the $\Phi _{j}$ are $q_{0}$ and the $2N+1$ parameters $\varphi
_{\alpha }$ on the surface ${\cal D}$ (see \cite{CH}), then we will also
find a local solution of the partial-differential equation (\ref{Ap.1}).

b.- Let us now consider eq.(\ref{1.3}) from a physical point of view, in
order to see if the Lipschitz conditions are satisfied. Introducing the time 
$t$ $=q_{0}$ and, therefore, $p_{0}=H$ independent of $t$, eqs.(\ref{Ap.3})
read 
\begin{equation}
\frac{dq_{i}}{dt}=\frac{\partial H(q,p)}{\partial p_{i}},\quad \frac{dp_{i}}{%
dt}=-\frac{\partial H(q,p)}{\partial q_{i}}  \label{Ap.5}
\end{equation}
namely, the Hamilton equations. This system is Lipschitzian if there is a $%
M>0$ such that 
\[
\left| \frac{\partial H(Q,P)}{\partial p_{i}}-\frac{\partial H(q,p)}{%
\partial p_{i}}\right| <M\left( \sum_{i}|Q_{i}-q_{i}|+|P_{i}-p_{i}|\right) 
\]
\begin{equation}
\left| \frac{\partial H(Q,P)}{\partial q_{i}}-\frac{\partial H(q,p)}{%
\partial q_{i}}\right| <M\left( \sum_{i}|Q_{i}-q_{i}|+|P_{i}-p_{i}|\right)
\label{Ap.6}
\end{equation}
in a certain domain ${\cal D}$ of the phase space. These conditions are
fulfilled if the derivatives $\frac{\partial ^{2}H(q,p)}{\partial
q_{i}\partial q_{j}},\frac{\partial ^{2}H(q,p)}{\partial q_{i}\partial p_{j}}%
,\frac{\partial ^{2}H(q,p)}{\partial p_{i}\partial p_{j}}$ are bounded in
the domain ${\cal D}$ (see e.g. \cite{dela} p.141). We can understand the
meaning of this requirement in a physical example. Let us consider the
Hamiltonian 
\begin{equation}
H(q_{1},q_{2},p_{1},p_{2})=\frac{p_{1}^{2}}{2}+\frac{q_{1}^{2}}{2}+\frac{%
p_{2}^{2}}{2}+\frac{q_{2}^{2}}{2}+V(q_{1},q_{2})  \label{Ap.7}
\end{equation}
For this particular case, the derivatives of the potential $\frac{\partial
^{2}V(q,p)}{\partial q_{i}\partial q_{j}}$ must be bounded in ${\cal D}.$
Let us consider the potentials $V(q_{1},q_{2})=(q_{1}-q_{2})^{\alpha }$. For 
$\alpha \geq 2$, these potentials have their $\frac{\partial ^{2}V(q,p)}{%
\partial q_{i}\partial q_{j}}$ bounded in ${\cal D}$ and the system is
Lipschitzian (this is the case in quadratic or polynomial potentials used in
many models). For $\alpha <2$, these potentials have their $\frac{\partial
^{2}V(q,p)}{\partial q_{i}\partial q_{j}}$ not bounded in ${\cal D}$ and the
system is not Lipschitzian (this is the case of Newton potential $%
(q_{1}-q_{2})^{-1}$). But realistic physical potentials are always bounded,
e.g. the Lenard-Jones potential for finite energies. In fact, there is
always a central repulsive core. This is not the case of the (mathematical)
three bodies problem with Newton potentials, provided that all the planets
can get infinitely close; but, of course, this is not a physical situation.
In conclusion, all the physical potentials are Lipschitzian.

c.- Let us now consider the solution of eq.(\ref{M.2}) in a powers of $\hbar 
$ expansion. A simple computation shows that, in this case, the problem
reduces, for each term, to the one studied in paragraphs a.- and b.- of this
Appendix, but for non-homogeneous equations. In fact,

\[
\{O_{I}(\phi ),O_{J}(\phi )\}_{mb}=\{O_{I}(\phi ),O_{J}(\phi )\}_{pb}+\hbar
^{2}A^{(1)}(\phi )+...=0 
\]
Then, if 
\[
O_{I}(\phi )=O_{I}^{(0)}(\phi )+\hbar ^{2}O_{I}^{(1)}(\phi )+... 
\]
we have that 
\[
\{O_{I}^{(0)}(\phi ),O_{J}^{(0)}(\phi )\}_{pb}+\hbar ^{2}\left(
\{O_{I}^{(1)}(\phi ),O_{J}^{(1)}(\phi )\}_{pb}+A^{(1)}(\phi )\right) +...=0 
\]
so, 
\[
\{O_{I}^{(0)}(\phi ),O_{J}^{(0)}(\phi )\}_{pb}=0, 
\]
\[
\{O_{I}^{(1)}(\phi ),O_{J}^{(1)}(\phi )\}_{pb}+A^{(1)}(\phi )=0,... 
\]
namely, the non-homogeneous version of eq.(\ref{Ap.1}), which can be solved
by means of the same method. In this case, the domain of the solution will
be the intersection of the domains of solution of all these equations.

\section{Sinai Billiard.}

Let us consider the Sinai billiard of fig.1 \cite{Stockmann}. It is clear
that, when the is confined to the inside of the billiard, the trajectories
are defined by two independent constants of the motion, $H$ and $P_{x}$ (or $%
H$ and $P_{y}$, or $P_{x}$ and $P_{y}$), which constitute a complete set of
local (i.e. in the interior $D_{0}$ of the billiard) constants of the motion
in involution. When the ball strikes the boundaries, it is symmetrically
reflected, i.e. the incident angle is equal to the reflected angle, and the
value of some of the constants of the motion changes: for the two horizontal
boundaries, $H$ and $P_{x}$ still constitute a complete set of local
constants of the motion in involution, but $P_{y}$ changes its sign; for the
vertical boundary, $H$ and $P_{y}$ still constitute a complete set of local
constants of the motion in involution, but $P_{x}$ changes its sign.

Without modifying the physical characterization of the example, we can
replace the rigid walls with infinitely height potential barriers of width $%
d $, namely, the potentials $V(x),V(y)$ and $V(r)$ of figure 2. (e.g. $V(x)$
behaves as $V(0)=0,V^{\prime }(0)=0,V(-d)\rightarrow \infty $)$.$ Due to the
symmetry of the potentials (translation symmetry for $V(x)$ and $V(y),$
rotation symmetry for $V(r)$), the reflections are still symmetric, i.e. the
ball climbs the potential walls and then falls down with symmetrical motion.
Calling $D_{1}$ and $D_{3}$ the domains in the potential of the $x$ walls, $%
D_{2}$ that of the $y$ wall, and $D_{4}$ that of the curved wall, we see
that $x$ is a cyclic variable in $D_{1}$ and $D_{3}$, $y$ is a cyclic
variable in $D_{2}$, and $\theta $ is a cyclic angular variable in $D_{4%
\text{ }}$. Therefore, we have the following local constants of the motion
in each domain:

\qquad \qquad \qquad \qquad \qquad $D_{0}:\qquad H\qquad P_{x}\qquad ($or $%
P_{y})$

\qquad \qquad \qquad \qquad \qquad $D_{1}:\qquad H\qquad P_{x}\qquad $

\qquad \qquad \qquad \qquad \qquad $D_{2}:\qquad H\qquad P_{y}\qquad $

\qquad \qquad \qquad \qquad \qquad $D_{3}:\qquad H\qquad P_{x}\qquad $

\qquad \qquad \qquad \qquad \qquad $D_{4}:\qquad H\qquad P_{\theta }\qquad $

In summary, we have found five domains, each one with two constants of the
motion in involution. If $d\rightarrow 0$, we go from fig.2 to fig.1.

{\bf Figure Caption:}

fig.1 A Sinai billiard.

fig.2 A Sinai billiard with potential barriers.

\end{document}